\documentclass[12pt]{article}
\usepackage{epsfig}

\renewcommand{\thefootnote}{\fnsymbol{footnote}}

\begin{document}

\title{
\begin{flushright}
\begin{minipage}{0.2\linewidth}
\normalsize
%arXiv:YYMM.NNNN \\
TU-834 \\
KUNS-2173 \\*[50pt]
\end{minipage}
\end{flushright}
{\Large \bf 
Three generation magnetized orbifold models
\\*[20pt]}}

\author{Hiroyuki~Abe$^{1,}$\footnote{
E-mail address: abe@tuhep.phys.tohoku.ac.jp}, \ 
Kang-Sin~Choi$^{2,}$\footnote{
E-mail address: kschoi@gauge.scphys.kyoto-u.ac.jp}, \ 
Tatsuo~Kobayashi$^{2,}$\footnote{
E-mail address: kobayash@gauge.scphys.kyoto-u.ac.jp} \ \\ and \ 
Hiroshi~Ohki$^{2,}$\footnote{E-mail address: ohki@scphys.kyoto-u.ac.jp
}\\*[20pt]
$^1${\it \normalsize 
Department of Physics, Tohoku University, 
Sendai 980-8578, Japan} \\
$^2${\it \normalsize 
Department of Physics, Kyoto University, 
Kyoto 606-8502, Japan} \\*[50pt]}

\date{
\centerline{\small \bf Abstract}
\begin{minipage}{0.9\linewidth}
\medskip 
\medskip 
\small
We study three generation models in the four-dimensional 
spacetime, which can be derived from the ten-dimensional 
${\cal N}=1$ super Yang-Mills theory on the orbifold background 
with a non-vanishing magnetic flux. We classify the flavor 
structures and show possible patterns of Yukawa matrices. 
Some examples of numerical studies are also shown. 
\end{minipage}
}

\begin{titlepage}
\maketitle
\thispagestyle{empty}
\clearpage
%\tableofcontents
\thispagestyle{empty}
\end{titlepage}

\renewcommand{\thefootnote}{\arabic{footnote}}
\setcounter{footnote}{0}

\section{Introduction}

Extra dimensional field theories, in particular 
string-derived extra dimensional field theories, 
play important roles in particle physics, e.g. 
as an origin of the flavor structure 
including the hierarchy of quark/lepton masses 
and mixing angles.
How to derive chiral theory is a key issue 
when our starting point is extra dimensional theory.
Introduction of a magnetic flux in extra dimensional space 
is one of interesting ways to obtain chiral theory.
Indeed, several studies have been carried out on models 
with magnetic fluxes 
in field theories and superstring theories~\cite{Manton:1981es,
Witten:1984dg,Bachas:1995ik,Berkooz:1996km,Blumenhagen:2000wh,
Angelantonj:2000hi,Cremades:2004wa,Troost:1999xn}.
Furthermore, magnetized D-brane models are 
T-duals of intersecting D-brane models 
and within the latter framework several interesting 
models have been constructed~\cite{Berkooz:1996km,Blumenhagen:2000wh,
Angelantonj:2000hi,Aldazabal:2000dg,
Blumenhagen:2000ea,Cvetic:2001tj}\footnote{
See for a review \cite{Blumenhagen:2005mu} and references therein.}.

Zero-modes are quasi-localized on the torus with the magnetic flux.
The number of zero-modes, which corresponds to 
the generation number, is determined by 
the value of the magnetic flux in the same way as 
that the generation number is determined by the 
intersecting number in intersecting D-brane models.
Yukawa couplings among zero-modes in four-dimensional effective theory 
are obtained by overlap integral of zero-mode 
profiles in extra dimensions.
Large or suppressed Yukawa couplings can be 
derived depending on the size of overlap integral.
That is, when zero-modes are quasi-localized 
far away from each other, their couplings in 4D 
effective field theory are suppressed.
On the other hand, when their localized points are close 
to each other, 4D effective Yukawa couplings would be 
of ${\cal O}(1)$.
Thus, magnetized torus models are quite interesting 
to derive realistic models, in particular 
a realistic flavor structure.
However, it is still a challenging issue 
to derive realistic mass matrices of quarks and leptons.

Orbifolding the extra dimensions is another way to derive 
chiral theory \cite{Dixon:1985jw}.
In Ref.~\cite{Abe:2008fi}, magnetized orbifold models have been 
studied\footnote{Other geometrical backgrounds with a magnetic 
flux have also been studied \cite{Conlon:2008qi,Marchesano:2008rg}.}. 
Phenomenological aspects in magnetized orbifold models 
are different from those in magnetized torus models. 
Some of zero-modes are projected out by the orbifold 
projection.
However, odd modes as well as even modes could correspond to zero-modes, 
although odd modes correspond to only massive modes on 
orbifolds without the magnetic flux.
Then, the generation number is smaller than the number of 
the magnetic flux, i.e. one in magnetized torus models with 
the same magnetic flux. 
Thus, a new type of flavor structure can appear 
in magnetized orbifold models.
Hence, it is quite important to study in detail phenomenological 
aspects of magnetized orbifold models.
That is our purpose in this paper.
We classify three generation models and 
study predicted patterns of Yukawa matrices.

The paper is organized as follows.
In section 2, we give a review on magnetized orbifold models.
In section 3, we classify three generation models on 
the orbifold with magnetic fluxes.
In section 4, we study Yukawa couplings in three generation models 
and we show explicitly an example of 
numerical studies on our models.
Section 5 is devoted to conclusion and discussion.
In Appendix, we show explicitly all of possible 
Yukawa matrices 
in our three generation models.

\section{Magnetized extra dimensions}

Here, we give a review on extra dimensional models with 
a magnetic flux on torus and orbifold backgrounds 
\cite{Cremades:2004wa,Abe:2008fi}.

\subsection{$U(N)$ gauge theory on $(T^2)^3$}

We start with ${\cal N}=1$ ten-dimensional $U(N)$ super Yang-Mills theory.
We consider the background $R^{3,1}\times (T^2)^3$,  
whose coordinates are denoted by
$x_\mu$ $(\mu=0,\cdots, 3)$ for the uncompact space $R^{3,1}$
and $y_m$ $(m=4, \cdots, 9)$ for the compact space $(T^2)^3$.
At the first stage, we use orthogonal coordinates 
of the compact space  
and choose the torus metric such that $y_m$ is 
identified by $y_m+n_m$ with $n_m=$ integer, i.e. 
$y_m \sim y_m +1$.
At the end of this subsection, we will 
extend it by introducing  the complex structure.
Also, we can extend the following discussions 
to ${\cal N}=1$ super Yang-Mills theory on 
$R^{3,1}\times (T^2)^n$.

The Lagrangian is given by 
\begin{eqnarray}
{\cal L} &=& 
-\frac{1}{4g^2}{\rm Tr}\left( F^{MN}F_{MN}  \right) 
+\frac{i}{2g^2}{\rm Tr}\left(  \bar \lambda \Gamma^M D_M \lambda
\right),
\nonumber
\end{eqnarray}
where $M,N=0,\cdots, 9$.
Here, $\lambda$ denotes gaugino fields, $\Gamma^M$ is the 
gamma matrix for ten-dimensions and 
the covariant derivative $D_M$ is given as 
\begin{eqnarray}
D_M\lambda &=& \partial_M \lambda - i [A_M, \lambda],
\end{eqnarray}
where $A_M$ is the vector field.
Furthermore, the field strength $F_{MN}$ is given by 
\begin{eqnarray}
F_{MN} &=& \partial_M A_N - \partial_N A_M -i[A_M,A_N].
\end{eqnarray}

The gaugino fields $\lambda$ and the vector fields $A_m$ 
corresponding to the compact directions are decomposed as 
\begin{eqnarray}
\lambda(x,y) &=& \sum_n \chi_n(x) \otimes \psi_n(y), 
\nonumber \\
A_m(x,y) &=& \sum_n \varphi_{n,m}(x) \otimes \phi_{n,m}(y).
\nonumber
\end{eqnarray}
Here, we concentrate on zero-modes, $\psi_0(y)$ 
and we denote them as $\psi(y)$ 
by omitting the subscript ``0''.
Furthermore, the internal part $\psi(y)$ is 
decomposed as a product of the $i$-th $T^2$ parts, 
i.e. $\psi_{(i)}(y_{2i+2},y_{2i+3})$.
Each of  $\psi_{(i)}(y_{2i+2},y_{2i+3})$ is two-component 
spinor,  $\psi_{(i)}=(\psi_{(i)+},\psi_{(i)-})^T$, 
and their chirality for the $i$-th $T^2$ part is denoted by $s_i$.
We use the gamma matrix $\tilde \Gamma^m$ 
corresponding to the $i$-th $T^2$ as 
\begin{eqnarray}
\tilde \Gamma^{2i+2} 
&=& \left(
\begin{array}{cc}
0 & 1 \\
1 & 0 
\end{array}
\right), \qquad 
\tilde \Gamma^{2i+3} 
\ = \ \left(
\begin{array}{cc}
0 & -i \\
i & 0 
\end{array}
\right),
\end{eqnarray}
and the total gamma matrices are obtained as 
their direct products with the four-dimensional part.

Here, we introduce the magnetic flux in the background as
$F_{45}, F_{67}$ and  $F_{89}$, which are 
given by 
\begin{eqnarray}
F_{45} &=& 2 \pi \left(
\begin{array}{ccc}
M^{(1)}_1 {\bf 1}_{N_1\times N_1} & & 0  \\
 & \ddots & \\
0 & & M^{(1)}_n {\bf 1}_{N_n\times N_n}
\end{array}\right), 
\nonumber \\
F_{67} &=& 2 \pi\left(
\begin{array}{ccc}
M^{(2)}_1 {\bf 1}_{N_1\times N_1} & & 0 \\
 & \ddots & \\
0 & & M^{(2)}_n {\bf 1}_{N_n\times N_n}
\end{array}\right),  
\nonumber \\
F_{89} &=& 2 \pi \left(
\begin{array}{ccc}
M^{(3)}_1 {\bf 1}_{N_1\times N_1} & & 0 \\
 & \ddots & \\
0 & & M^{(3)}_n {\bf 1}_{N_n\times N_n}
\end{array}\right). 
\nonumber
\end{eqnarray}
This background breaks the gauge group $U(N)$ as 
$U(N) \rightarrow \prod_{a=1}^n U(N_a)$ with $N=\sum_a N_a$.
We concentrate on an Abelian flux, 
although in general non-Abelian magnetic fluxes, 
which reduce ranks of gauge groups, are possible 
\cite{'t Hooft:1979uj,Alfaro:2006is,vonGersdorff:2007uz}.

Here we focus on the $U(N_a) \times U(N_b)$ part, 
which has the magnetic flux, 
\begin{eqnarray}
F_{2i+2,2i+3} &=& 2 \pi\left(
\begin{array}{cc}
M^{(i)}_a {\bf 1}_{N_a\times N_a}  & 0 \\
0 & M^{(i)}_b {\bf 1}_{N_b\times N_b}
\end{array}\right),
\label{eq:6D-F-ab-block}
\end{eqnarray}
for $i=1,2,3$.
We use the following gauge,
\begin{eqnarray}
A_{2i+2} &=& 0, \qquad 
A_{2i+3} \ = \  F_{2i+2,2i+3}~y_{2i+2}.
\label{eq:gauge}
\end{eqnarray}
Similarly, the gaugino fields $\lambda$ and their 
$i$-th torus parts are 
decomposed as 
\begin{eqnarray}
\lambda(x,y) 
&=& \left(
\begin{array}{cc}
\lambda^{aa}(x,y)   & \lambda^{ab}(x,y)  \\
\lambda^{ba}(x,y) & \lambda^{bb}(x,y) 
\end{array}\right), \qquad 
\psi_{(i)}(y) 
\ = \ \left(
\begin{array}{cc}
\psi_{(i)}^{aa}(y)   & \psi_{(i)}^{ab}(y)  \\
\psi_{(i)}^{ba}(y) & \psi_{(i)}^{bb}(y) 
\end{array}\right).
\label{eq:gaugino-ab-block}
\end{eqnarray}
The fields $\lambda^{aa}$ and $\lambda^{bb}$ 
correspond to the gaugino fields under the 
unbroken gauge group $U(N_a) \times U(N_b)$.
On the other hand, $\lambda^{ab}$ and $\lambda^{ba}$ correspond 
to bi-fundamental matter fields,
 $(N_a, \bar N_b)$ and $(\bar N_a,N_b)$, under the unbroken 
gauge group $U(N_a) \times U(N_b)$.
The Dirac equations for these gaugino fields corresponding to 
zero-modes are obtained as 
\begin{eqnarray}
\left(
\begin{array}{cc}
\bar \partial_i \psi_{(i)+}^{aa} & [\bar  \partial_i +2\pi
(M^{(i)}_a-M^{(i)}_b)y_{2i+2} ] 
\psi^{ab}_{(i)+} \cr \cr
  [\bar \partial_i + 2\pi(M^{(i)}_b-M^{(i)}_a)y_{2i+2}] \psi^{b
a}_{(i)+} 
& \bar \partial_i \psi_{(i)+}^{bb}  \\
\end{array}
\right) &=& 0, 
\nonumber \\*[5pt] 
\left(
\begin{array}{cc}
\partial_i \psi_{(i)-}^{aa} & 
[\partial_i - 2\pi(M^{(i)}_a-M^{(i)}_b)y_{2i+2}] 
\psi^{ab}_{(i)-}  \cr \cr
[\partial_i -2 \pi(M^{(i)}_b-M^{(i)}_a)y_{2i+2}] \psi^{ba}_{(i)
-} & 
\partial_i \psi_{(i)-}^{bb} \\
\end{array}
\right) &=& 0, 
\nonumber
\end{eqnarray}
where $\bar \partial_i = \partial_{2i+2}  +i \partial_{2i+3}$ and 
$\partial_i = \partial_{2i+2} -i \partial_{2i+3}$.
The gaugino fields, $\psi^{aa}$ and $\psi^{bb}$, for 
the unbroken gauge symmetry have no effect from the magnetic 
flux in their Dirac equations.
Hence, they have the same zero-modes as those on 
$(T^{2})^3$ without the magnetic flux.
On the other hand, the magnetic flux appears in 
the zero-mode equations of $\psi^{ab}$ and $\psi^{ba}$ 
corresponding  to bi-fundamental matter fields,
 $(N_a, \bar N_b)$ and $(\bar N_a,N_b)$.
Furthermore, they satisfy the following boundary conditions,
\begin{eqnarray}
\psi^{ab}_{s_i}(y_{2i+2}+1,y_{2i+3}) &=& 
e^{2\pi i s_i(M^{(i)}_a-M^{(i)}_b)y_{2i+3}}
\psi^{ab}_{s_i}(y_{2i+2},y_{2i+3}), 
\nonumber \\
\psi^{ba}_{s_i}(y_{2i+2}+1,y_{2i+3}) &=& 
e^{2\pi i s_i(M^{(i)}_b-M^{(i)}_a)y_{2i+3}}
\psi^{ba}_{s_i}(y_{2i+2},y_{2i+3}),  
\nonumber \\
\psi^{ab}_{s_i}(y_{2i+2},y_{2i+3}+1) &=&
\psi^{ab}_{s_i}(y_{2i+2},y_{2i+3}), 
\nonumber \\ 
\psi^{ba}_{s_i}(y_{2i+2},y_{2i+3}+1) &=& 
\psi^{ba}_{s_i}(y_{2i+2},y_{2i+3}), 
\nonumber 
\nonumber
\end{eqnarray}
because of Eq.~(\ref{eq:gauge}).

For the $i$-th $T^2$ with $M^{(i)}_a -M^{(i)}_b >0$,
the fields $\psi^{ab}_{(i)+}$ and $\psi^{ba}_{(i)-}$ have 
$|M^{(i)}_a - M^{(i)}_b |$ normalizable zero-modes, 
while $\psi^{ab}_{(i)-}$ and $\psi^{ba}_{(i)+}$ have 
no normalizable zero-modes.
Thus, we can derive chiral theory.
When $M^{(i)}_a -M^{(i)}_b <0$, 
$\psi^{ab}_{(i)-}$ and $\psi^{ba}_{(i)+}$ have 
$|M^{(i)}_a - M^{(i)}_b |$ normalizable zero-modes.
The normalizable wavefunction for the $j$-th zero mode 
is obtained as 
\begin{eqnarray}
\Theta^j(y_{2i+2},y_{2i+3}) 
&=& N_je^{-M\pi y_{2i+2}^2}\vartheta \left[
\begin{array}{c}
j/M \\ 0
\end{array} \right]\left( M(y_{2i+2}+iy_{2i+3}), Mi\right), 
\nonumber
\end{eqnarray}
for $M=|M^{(i)}_a - M^{(i)}_b|$ and 
$j=0,1,\cdots, N-1$, where $N_j$ is a normalization constant and 
\begin{eqnarray}
\vartheta \left[ 
\begin{array}{c}
j/M \\ 0
\end{array} \right]
\left( M(y_{2i+2}+iy_{2i+3}), Mi \right) 
&=& 
\sum_n e^{-M\pi (n+j/M)^2 +2\pi i (n+j/M)M(y_{2i+2}+iy_{2i+3})}, 
\nonumber
\end{eqnarray}
that is, the Jacobi theta-function.
Furthermore, we can introduce the complex structure modulus 
$\tau$ by replacing 
the above Jacobi theta-function as 
\begin{eqnarray}
\vartheta \left[
\begin{array}{c}
j/M \\ 0
\end{array} \right]
\left( M(y_{2i+2}+iy_{2i+3}), Mi\right) 
&\rightarrow& 
\vartheta \left[
\begin{array}{c}
j/M \\ 0
\end{array} \right]
\left( M(y_{2i+2}+\tau y_{2i+3}), M \tau \right).
\nonumber
\end{eqnarray}

The total number of bi-fundamental zero-modes 
is given by $\prod_{i=1}^3|M_a^{(i)}-M_b^{(i)}|$ and 
all of them have the same six-dimensional chirality 
${\rm sign} \left[ \prod_{i=1}^3(M_a^{(i)}-M_b^{(i)})\right]$.
Since the ten-dimensional chirality of gaugino fields is fixed, 
bi-fundamental zero-modes for either $(N_a,\bar N_b)$ or 
$(\bar N_a,N_b)$ appear for a fixed four-dimensional chirality.
That is, the total number of bi-fundamental zero-modes 
for $(N_a,\bar N_b)$ is equal to 
\begin{eqnarray}
I_{ab} &=& \prod_{i=1}^3(M_a^{(i)}-M_b^{(i)}).
\nonumber
\end{eqnarray}
When $I_{ab} <0$, this means 
that there appear $|I_{ab}|$ independent zero modes for 
$(\bar N_a,N_b)$.
It is also convenient to introduce the notation, 
$I^i_{ab} \equiv M_a^{(i)}-M_b^{(i)}$.
Zero-mode wavefunctions are given by a product
of two-dimensional parts, i.e. 
\begin{eqnarray}
\Theta^{i_1,i_2,i_3}(y) 
&=& \Theta^{i_1}(y_4,y_5) 
\Theta^{i_2}(y_6,y_7) \Theta^{i_3}(y_8,y_9),
\nonumber
\end{eqnarray}
for $i_1 =0,\cdots, (|M_a^{(1)}-M_b^{(1)}|-1)$, 
$i_2 =0,\cdots, (|M_a^{(2)}-M_b^{(2)}|-1)$ and 
$i_3 = 0,\cdots, (|M_a^{(3)}-M_b^{(3)}|-1)$.

\subsection{$U(N)$ gauge theory on magnetized orbifolds 
$T^6/(Z_2\times Z'_2)$}

Here we review on the $T^6/(Z_2\times Z'_2)$ 
orbifold with a magnetic flux \cite{Abe:2008fi}.

\subsubsection{$T^2/Z_2$ orbifold}

First, let us study the $U(N)$ gauge theory on the orbifold $T^2/Z_2$ 
with the coordinates $(y_4,y_5)$, 
which transform as 
\begin{eqnarray}
y_4 &\rightarrow& -y_4, \qquad y_5 \rightarrow -y_5,
\nonumber
\end{eqnarray}
under the $Z_2$ orbifold twist.
Here, we associate the $Z_2$ twist with the $Z_2$ 
action in the gauge space as 
\begin{eqnarray}
A_\mu(x,-y) 
&=& P A_\mu(x,y)P^{-1}, \qquad A_m(x,y) = -P A_m (x,y)P^{-1},
\nonumber
\end{eqnarray}
and the $Z_2$ boundary conditions for 
gaugino fields,
\begin{eqnarray}
\lambda_{\pm}(x,-y) &=& \pm P \lambda_{\pm}(x,y) P^{-1},
\nonumber
\end{eqnarray}
where the $Z_2$ projection $P$ must 
satisfy $P^2 =1$.

We focus on the $U(N_a) \times U(N_b)$ 
block (\ref{eq:6D-F-ab-block}), (\ref{eq:gaugino-ab-block}) and 
consider the spinor fields, $\lambda^{aa}_\pm$, 
$\lambda^{ab}_\pm$, $\lambda^{ba}_\pm$ and $\lambda^{bb}_\pm$,
in particular bi-fundamental fields 
$\lambda^{ab}_\pm$ and $\lambda^{ba}_\pm$,
where $\pm$ denotes the chirality $s_i$ in the extra dimension.
Without the $Z_2$ projection, there are $|M_a - M_b|$ 
zero modes for $\lambda^{ab}_\pm$ and $\lambda^{ba}_\pm$.
For example, when $M_a - M_b>0$, $\lambda^{ab}_+$ as well as 
$\lambda^{ba}_-$ has  $(M_a - M_b)$ zero modes with 
the wavefunctions $\Theta^j$ for $j=0, \cdots, (M_a - M_b -1)$.
When we consider the $Z_2$ projection, either even or odd modes 
of them remain.
Here note that 
\begin{eqnarray}
\Theta^j(-y_4,-y_5) &=& \Theta^{M-j}(y_4,y_5),
\nonumber
\end{eqnarray}
where $\Theta^{M}(y_4,y_5)=\Theta^{0}(y_4,y_5)$.
That is, even and odd functions are given by 
\begin{eqnarray}
\Theta^j_{\rm even} 
&=& \frac{1}{\sqrt 2}(\Theta^j + \Theta^{M-j}), 
\nonumber \\
\Theta^j_{\rm odd} 
&=& \frac{1}{\sqrt 2}(\Theta^j - \Theta^{M-j}), 
\label{eq:even-odd-function}
\end{eqnarray}
respectively.
For example, when we consider the projection $P$ such that 
$\lambda^{ab}_+(x,-y) =\lambda^{ab}_+(x,y)$,  
only zero-modes corresponding to $\Theta^j_{\rm even}$ remain 
and the number of zero-modes is equal to $(M_a - M_b)/2 + 1$ for 
$(M_a - M_b)=$ even and $(M_a - M_b+1)/2 $ for $(M_a - M_b)=$ odd.
On the other hand, when we consider the projection $P$ such that 
$\lambda^{ab}_+(x,-y) =-\lambda^{ab}_+(x,y)$,  
only zero-modes corresponding to $\Theta^j_{\rm odd}$ remain 
and the number of zero-modes is equal to $(M_a - M_b)/2 - 1$ for 
$(M_a - M_b)=$ even and $(M_a - M_b - 1)/2 $ for $(M_a - M_b)=$ odd.
The same holds true for $\lambda^{ba}_-$.
Table 1 shows the numbers of zero-modes with even 
and odd wavefunctions for $M \leq 10$.
\begin{table}[t]
\begin{center}
\begin{tabular}{|c|ccccccccccc|}\hline
$M$ & 0 & 1 & 2 & 3 & 4& 5 & 6 & 7 & 8 & 9& 10 
\\ \hline 
even & 1 & 1 & 2 & 2 & 3 & 3 & 4 & 4 & 5 & 5 & 6 
\\ \hline
odd & 0 & 0 & 0& 1 & 1 & 2 & 2 & 3 & 3 & 4 & 4 
\\ \hline
\end{tabular}
\end{center}
\caption{The numbers of zero-modes for even and odd wavefunctions.}
\label{even-odd-zero-modes}
\end{table}

\subsubsection{$T^6/(Z_2 \times Z_2')$}

Here, we can extend the previous analysis on 
the two-dimensional orbifold $T^2/Z_2$ to 
the $U(N)$ gauge theory on the six-dimensional 
orbifold $T^6/(Z_2\times Z'_2)$.
We consider two independent twists, $Z_2$ and $Z_2'$.
The $Z_2$ twist acts on the six-dimensional coordinates 
$y_m$ ($m=4,\cdots, 9$) as
\begin{eqnarray}
y_m &\rightarrow& -y_m~~({\rm for~~}m=4,5,6,7), \qquad 
y_n \ \rightarrow \ y_n~~({\rm for~~}n=8,9), 
\nonumber
\end{eqnarray}
and the $Z_2'$ twist acts as 
\begin{eqnarray}
y_m &\rightarrow& -y_m~~({\rm for~~}m=4,5,8,9), \qquad 
y_n \ \rightarrow \ y_n~~({\rm for~~}n=6,7).
\nonumber
\end{eqnarray}
If the magnetic flux is vanishing, we realize 
four-dimensional ${\cal N}=1$ supersymmetric gauge 
theories for the orbifold 
$T^6/(Z_2\times Z'_2)$. 
The bi-fundamental matter fields 
$\lambda^{ab}_{s_1,s_2,s_3}$, $\lambda^{ba}_{s_1,s_2,s_3}$
with the chirality $s_i$ corresponding 
to the $i$-th $T^2$ are also introduced.
Their $Z_2$ boundary conditions are given by 
\begin{eqnarray}
\lambda_{s_1,s_2,s_3}(x,-y_m,y_n) 
&=& s_1s_2P\lambda_{s_1,s_2,s_3}(x,y_m,y_n) P^{-1},
\nonumber
\end{eqnarray}
with $m=4,5,6,7$ and $n=8,9$ for $\lambda^{aa}_{s_1,s_2,s_3}$, 
$\lambda^{ab}_{s_1,s_2,s_3}$, $\lambda^{ba}_{s_1,s_2,s_3}$ 
and $\lambda^{bb}_{s_1,s_2,s_3}$. 
Similarly, the $Z'_2$ boundary conditions are given by 
\begin{eqnarray}
\lambda_{s_1,s_2,s_3}(x,-y_m,y_n) 
&=& s_1s_3P'\lambda_{s_1,s_2,s_3}(x,y_m,y_n) P'^{-1},
\nonumber
\end{eqnarray}
with $m=4,5,8,9$ and $n=6,7$. 
Then, depending on the projections $P$ and $P'$, 
even or odd modes for the $i$-th torus remain 
such as $\Theta^{j,M}_{\rm even}$ or $\Theta^{j,M}_{\rm odd}$.
Their products such as 
$\prod_{i=3}^3\Theta^{j_i,M}_{\rm even,odd}(y_{2i+2},y_{2i+3})$
provide with zero-modes on the $T^6/(Z_2 \times Z'_2)$.

\section{Three generation magnetized orbifold models}

In this section, 
we consider the $U(N_a)\times U(N_b)\times U(N_c)$ models,
which lead to three 
families of bi-fundamental matter fields, 
$(N_a,\bar N_b)$ and $(\bar N_a,N_c)$.
Such a gauge group is derived by starting with the $U(N)$ group 
and introducing the following form of the magnetic flux,
\begin{eqnarray}
F_{45} &=& 2 \pi \left(
\begin{array}{ccc}
M^{(1)}_a {\bf 1}_{N_a\times N_a} & & 0  \\
 & M^{(1)}_b {\bf 1}_{N_b\times N_b} & \\
0 & & M^{(1)}_c {\bf 1}_{N_c\times N_c}
\end{array}\right),  
\nonumber \\
F_{67} &=& 2 \pi\left(
\begin{array}{ccc}
M^{(2)}_a {\bf 1}_{N_a\times N_a} & & 0 \\
 & M^{(2)}_b {\bf 1}_{N_b\times N_b} & \\
0 & & M^{(2)}_c {\bf 1}_{N_c\times N_c}
\end{array}\right),  
\nonumber \\
F_{89} &=& 2 \pi \left(
\begin{array}{ccc}
M^{(3)}_a {\bf 1}_{N_a\times N_a} & & 0 \\
 & M^{(3)}_b {\bf 1}_{N_b\times N_b} & \\
0 & & M^{(3)}_c {\bf 1}_{N_c\times N_c}
\end{array}\right), 
\nonumber
\end{eqnarray}
where $N=N_a+N_b+N_c$.
For $N_a=4, N_b=2$ and $N_c=2$, we can realize 
the Pati-Salam gauge group up to $U(1)$ factors, 
some of which may be anomalous and become massive 
by the Green-Schwarz mechanism.
Then, the bi-fundamental matter fields, 
$(N_a,\bar N_b)$ and $(\bar N_a,N_c)$ correspond 
to left-handed and right-handed matter fields.
In addition, the bi-fundamental matter fields 
$(N_b,\bar N_c)$ correspond to higgsino fields.
We assume that supersymmetry is preserved 
at least locally at the $a-b$ sector, $b-c$ sector 
and $c-a$ sector.\footnote{See for the supersymmetric conditions 
e.g. Ref.~\cite{Cremades:2004wa,Troost:1999xn}.}
Then, the number of Higgs scalar fields are the same 
as the number of higgsino fields.
There are no tachyonic modes at the tree level.
Indeed, in intersecting D-brane models 
it would be one of convenient ways towards  
realistic models to derive the Pati-Salam model at some stage 
and to break the gauge group to 
the group $SU(3) \times SU(2)_L \times U(1)$. 
(See e.g. Ref.~\cite{Cvetic:2001tj,Blumenhagen:2002gw} and references 
therein.)\footnote{
See for the Pati-Salam model in heterotic orbifold 
models e.g. Ref.~\cite{Kobayashi:2004ud}, 
where $SU(4)\times SU(2)_L \times SU(2)_R$ 
is broken to the standard gauge group by vacuum expectation values 
of scalar fields, $(4,1,2)$ and $(\bar 4,1,2)$, 
while in the intersecting D-brane models 
$SU(4)\times SU(2)_L \times SU(2)_R$ is broken by splitting 
D-branes, that is, vacuum expectation values of adjoint scalar 
fields.}
At the end of this section, we give a comment on 
breaking of $SU(4) \times SU(2)_L \times SU(2)_R$
to $SU(3) \times SU(2)_L \times U(1)$.

In both cases with and without orbifolding, 
the total number of chiral matter fields is a product 
of the numbers of zero-modes corresponding to the 
$i$-th $T^2$ for $i=1,2,3$.
That is, the three generations are realized in the 
models, where the $i$-th $T^2$ has three 
zero-modes while each of the other tori has 
a single zero mode.
Thus, there are two types of flavor structures.
That is, in one type the three zero-modes 
corresponding to both left-handed matter fields 
$(N_a,\bar N_b)$ and right-handed matter fields 
$(\bar N_a, N_c)$ appear in the same $i$-th $T^2$, 
while each of the other tori has a single zero-mode 
for $(N_a,\bar N_b)$ as well as $(\bar N_a,N_c)$.
In the other type, three zero-modes of 
 $(N_a,\bar N_b)$ and $(\bar N_a,N_c)$ are 
originated from different tori.
The Yukawa coupling for 4D effective field theory 
is evaluated by the following overlap integral 
of zero-mode wavefunctions \cite{Green:1987mn}
\begin{eqnarray}
Y_{ij} &=& \int d^6y \psi_{Li}(y) \psi_{Rj}(y)  \phi_H(y), 
\nonumber
\end{eqnarray}
where $\psi_L(y)$, $\psi_R(y)$ and $\phi_H(y)$ 
denote zero-mode wave-functions of 
the left-handed, right-handed matter fields 
and Higgs field, respectively.
Note that the integral corresponding to each torus is 
factorized in the Yukawa coupling.
In the second type of flavor structure, 
one obtains the following 
form of Yukawa matrices,
\begin{eqnarray}
Y_{ij} &=& a_i b_j,
\nonumber
\end{eqnarray}
at the tree-level, because the flavor structure 
of left-handed and right-handed matter fields are 
originated from different tori.
This matrix, $Y_{ij}$, has rank one 
and that is not phenomenologically interesting, 
unless certain corrections appear.
Hence, we concentrate on the first type 
of the flavor structure.
In the first type, the flavor structure is 
originated from the single torus, where 
both three zero-modes of $(N_a,\bar N_b)$ 
and $(\bar N_a,N_c)$ appear.
We assign this torus with the first torus. 
On the other hand, 
the other tori, the second and third tori,  
do not lead to flavor-dependent aspects.
That is, Yukawa matrices are obtained 
as the following form,
\begin{eqnarray}
Y_{ij} &=& a^{(2)}a^{(3)}a^{(1)}_{ij},
\nonumber
\end{eqnarray}
where the structure of $a^{(1)}_{ij}$ is determined 
by only the first torus corresponding to three 
zero-modes $(N_a,\bar N_b)$ and $(\bar N_a,N_c)$ 
while the other tori contribute to overall factors 
$a^{(2)}$ and $a^{(3)}$.
Thus, we concentrate on the single torus, 
where both of three zero-modes $(N_a,\bar N_b)$ 
and $(\bar N_a,N_c)$ appear, i.e. the first torus.

\begin{table}[t]
\begin{center}
\begin{tabular}{c|ccc} 
    &  $\lambda^{ab}$ 
    &  $\lambda^{ca}$ 
    &  $\lambda^{bc}$ 
\\ \hline 
I   &  even & even & even  \\
II  &  even & odd  & odd   \\
II' &  odd  & even & odd   \\
III &  odd  & odd  & even  \\
\end{tabular}
\end{center}
\caption{Possible patterns of wavefunctions 
with non-vanishing Yukawa couplings 
for the first torus.}
\label{class}
\end{table}

Zero-mode wavefunctions are classified into even and odd modes 
under the $Z_2$ twist.
Only even or odd modes remain through 
the orbifold projection.
Furthermore, the 4D Yukawa couplings are 
non-vanishing for combinations among 
(even, even, even) wavefunctions and 
(even, odd, odd) wavefunctions, while Yukawa couplings 
vanish for combinations among 
(even, even, odd) wavefunctions and 
(odd, odd, odd) wavefunctions.
Thus, we study only the former case
with non-vanishing Yukawa couplings, 
that is, the combinations among 
(even, even, even) wavefunctions and 
(even, odd, odd) wavefunctions.
Hence, we are interested in four types of 
combinations of wavefunctions for the first torus, 
as shown in Table \ref{class}.
The II' type of combinations is obtained by 
exchanging the left and right-handed matter fields 
in the II type.
Thus, we study explicitly the three types, I, II and III.

We can realize three even zero-modes when $|I^{(1)}_{ab}|=4,5$,
as shown in Table \ref{even-odd-zero-modes}.
On the other hand, three odd zero-modes can appear 
when $|I^{(1)}_{ab}|=7,8$.
Furthermore, the consistency condition on magnetic fluxes 
requires 
\begin{eqnarray}
|I_{bc}^{(1)}| &=& |I_{ab}^{(1)}|\pm |I_{ca}^{(1)}|.
\nonumber
\end{eqnarray}
Thus, the number of Higgs and higgsino fields are 
constrained.
Table \ref{higgs} shows all of possible magnetic fluxes 
for the three types, I, II and III.
The fourth and fifth columns of the table show 
possible sizes of magnetic fluxes for $|I_{bc}^{(1)}|$ 
and the number of zero-modes corresponding to 
the Higgs fields.
As a result, flavor structures of our models 
with Yukawa couplings are classified into 20 classes.
However, the model with 
$(|I_{ab}^{(1)}|,|I_{ca}^{(1)}|,|I_{bc}^{(1)}|) = (5,7,2)$ 
has no zero-modes for the Higgs fields.
Thus, we do not consider this case, 
but we will study the other 19 classes 
in Table \ref{higgs}.
Therefore, we study possible flavor structures explicitly 
by deriving the coupling selection rule 
and evaluating values of Yukawa couplings in 
these 19 classes.
That is the purpose of the next section.

\begin{table}[t]
\begin{center}
\begin{tabular}{c|ccc|c} 
    &  $|I_{ab}^{(1)}|$ 
    &  $|I_{ca}^{(1)}|$ 
    &  $|I_{bc}^{(1)}|$ 
    &  the numbers of \\ 
    &  &  &  & Higgs zero modes \\ \hline 
I   &  4 & 4 & 8  & 5  \\
    &  4 & 4 & 0  & 1  \\
    &  4 & 5 & 9  & 5  \\
    &  4 & 5 & 1  & 5  \\
    &  5 & 5 & 10 & 6  \\
    &  5 & 5 & 0  & 1  \\ \hline
II  &  4 & 7 & 11 & 5  \\
    &  4 & 7 & 3  & 1  \\
    &  4 & 8 & 12 & 5  \\
    &  4 & 8 & 4  & 1  \\
    &  5 & 7 & 12 & 5  \\
    &  5 & 7 & 2  & 0  \\
    &  5 & 8 & 13 & 6  \\
    &  5 & 8 & 3  & 1  \\ \hline
III &  7 & 7 & 14 & 8  \\
    &  7 & 7 & 0  & 1  \\
    &  7 & 8 & 15 & 8  \\
    &  7 & 8 & 1  & 1  \\
    &  8 & 8 & 16 & 9  \\
    &  8 & 8 & 0  & 1  \\
\end{tabular}
\end{center}
\caption{The number of Higgs fields of $(T^2)^1$ 
with non-vanishing Yukawa couplings.}
\label{higgs}
\end{table}

Before explicit study on flavor structures of 
19 classes in the next section, 
we give a comment on breaking of $SU(4)\times SU(2)_L \times SU(2)_R$.
At any rate, we need the $SU(3) \times SU(2)_L \times U(1)$ 
gauge group at low energy.
When the magnetic flux and orbifold projections lead to 
the $SU(4)\times SU(2)_L \times SU(2)_R$ gauge group from 
$U(8)$ as we have discussed so far, 
we need further breaking of $SU(4)\times SU(2)_L \times SU(2)_R$ 
to $SU(3) \times SU(2)_L \times U(1)$.
Such breaking can be realized by assuming 
non-vanishing vacuum expectation values (VEVs) 
of Higgs fields like adjoint scalar fields for 
$SU(4)$ and $SU(2)_R$ and/or 
bi-fundamental scalar fields like $(4,1,2)$ and 
$(\bar 4, 1,2)$ on fixed points.
Note that our models have degree of freedom to add any modes at 
the fixed points from the viewpoint of point particle field theory.
The above breaking may affect the structure of Yukawa matrices 
as higher dimensional operators.
However, we will show results on Yukawa matrices 
without such corrections.

Alternatively, magnetic fluxes and/or 
orbifold projections break $U(8)$ into 
$U(3)\times U(1)_1 \times U(2)_L \times U(1)_2 \times U(1)_3$.
The gauge group $U(3)\times U(1)_1$ would correspond to $U(4)$ and 
$U(1)_2 \times U(1)_3$ would correspond to $U(2)_R$. 
We assume that all the bi-fundamental matter fields under 
$U(3) \times U(1)_1$, i.e. extra colored modes, are projected out.
The bi-fundamental matter fields for 
$U(3) \times U(1)_2$ and $U(3) \times U(1)_3$ 
correspond to up and down sectors of right-handed quarks, respectively.
Similarly, up and down sectors of Higgs fields 
and right-handed charged leptons and neutrinos 
are obtained.
In this case, the classification of this section and 
patterns of Yukawa matrices, which will be studied 
in the next section and Appendix, are available for 
up-sector and down-sector quarks as well as the 
lepton sector.
However, the up sector and down sector can 
correspond to different classes of Table \ref{higgs}.
On the other hand, the up sector and 
down sector correspond to 
the same class in Table \ref{higgs}, when 
the  $SU(4)\times SU(2)_L \times SU(2)_R$ 
is broken by VEVs of Higgs fields 
on fixed points as discussed above.

\section{Yukawa couplings in three generation models}

\subsection{Yukawa interactions}

Following \cite{Cremades:2004wa,DiVecchia:2008tm}, 
first we show computation of Yukawa interactions 
on the torus with the magnetic flux.
Omitting the gauge structure and spinor structure,
the Yukawa coupling among left, right-handed matter fields 
and Higgs field corresponding to three zero-mode 
wavefunctions, $\Theta^{i,M_1}(z)$,  $\Theta^{j,M_2}(z)$ 
and $(\Theta^{k,M_3}(z))^*$, is written by
\begin{eqnarray}
Y_{ijk} &=& c \int dz d\bar{z}
\Theta^{i,M_1}(z) \Theta^{j,M_2}(z) (\Theta^{k,M_3}(z))^*,
\label{eq:yukawa}
\end{eqnarray}
where $z=x_4+\tau y_5$, 
$M_1 \equiv I^{(1)}_{ab}$, 
$M_2 \equiv I^{(1)}_{ca}$, 
$M_3 \equiv I^{(1)}_{cb}$ 
and $c$ is a flavor-independent contribution due to the other tori.
Note that $M_1+M_2=M_3$. 
Because of the gauge invariance, not the wavefunction 
$\Theta^{k,M_3}(z)$, but $(\Theta^{k,M_3}(z))^*$ 
appears in the Yukawa coupling~\cite{Cremades:2004wa}. 

By using the formula of the $\vartheta$ function, 
\begin{eqnarray}
\lefteqn{
\vartheta
\left[\begin{array}{c} r/N_1 \\ 0 \end{array} \right]
\left(z_1,N_1\tau \right)
\,\times\, 
\vartheta
\left[\begin{array}{c} s/N_2 \\ 0 \end{array} \right]
\left(z_2,N_2\tau \right)
}
\nonumber \\ &=& 
\sum_{m\in \mathcal{Z}_{N_1+N_2}}
\vartheta
\left[\begin{array}{c} \frac{r+s+N_1m}{N_1+N_2} \\ 0 \end{array} \right]
\left(z_1+z_2,\tau(N_1+N_2) \right) 
\nonumber \\ && \qquad\qquad \,\times\,
\vartheta
\left[\begin{array}{c} \frac{N_2r-N_1s+N_1N_2m}{N_1N_2(N_1+N_2)} \\ 0 
\end{array} \right]
\left(z_1N_2-z_2N_1,\tau N_1N_2(N_1+N_2) \right),
\nonumber
\end{eqnarray}
we can decompose $\Theta^{i,M_1}(z)\Theta^{j,M_2}(z)$ as 
\begin{eqnarray}
\Theta^{i,M_1}(z)\,\Theta^{j,M_2}(z) 
&=&\sum_{m\in \mathcal{Z}_{M_3}}\Theta^{i+j+M_1m,M_3}(z) 
\,\times\, 
\vartheta
\left[\begin{array}{c} 
\frac{M_2i-M_1j+M_1M_2m}{M_1M_2M_3} \\ 0
\end{array} \right]
\left(0,\tau M_1M_2M_3 \right). 
\nonumber
\end{eqnarray}
Wavefunctions satisfy the orthogonal condition
\begin{eqnarray}
\int dzd\bar{z}\,\Theta^{i,M}\,(\Theta^{j,M})^* &=& \delta_{ij}.
\nonumber
\end{eqnarray}
Then, the integral of three wavefunctions is represented by 
\begin{eqnarray}
Y_{ijk} 
&=& c \int dzd\bar{z}\,\Theta^{i,M_1}\,\Theta^{j,M_2}(\Theta^{k,M_3})^* 
\nonumber \\ &=& 
c \sum_{m=0}^{|M_3|-1} \vartheta 
\left[\begin{array}{c} 
\frac{M_2i-M_1j+M_1M_2m}{M_1M_2M_3} \\ 0
\end{array} \right]
\left(0,\tau M_1M_2M_3 \right) 
\times \delta_{i+j+M_1m,\,k+M_3 \ell},
\nonumber
\end{eqnarray}
where $\ell =$ integer.
%\begin{eqnarray}
%\delta_{i+j+M_1m,\,k\,{\rm mod}\,M_3} 
%&=& \left\{ 
%\begin{array}{ll}
%1 & \ (i+j+M_1m=k\ {\rm mod}\ M_3) \\*[5pt]
%0 & \ ({\rm others}) 
%\end{array}
%\right., 
%\nonumber
%\end{eqnarray}
Thus, we have the selection rule for allowed Yukawa
couplings as 
\begin{eqnarray}
i+j=k,
\nonumber
\end{eqnarray}
where $i, j$ and $k$ are defined up to mod 
$M_1, M_2$ and $M_3$, respectively.\footnote{
See for the selection rule in intersecting D-brane models, e.g. 
Ref.~\cite{Cremades:2003qj,Higaki:2005ie}.}
In addition, the Yukawa coupling $Y_{ijk}$, in particular 
its flavor-dependent part, is written by the $\vartheta$ function.
When $g.c.d.(M_1,M_3)=1$, a signle $\vartheta$ function appears 
in $Y_{ijk}$.
When $g.c.d.(M_1,M_3)=g \neq 1$, $g$ terms appear in $Y_{ijk}$ 
as 
\begin{eqnarray}
 Y_{ijk} = c \sum_{n=1}^g \vartheta \left[ \begin{array}{c}
    {M_2k - M_3j + M_2 M_3 \ell_0 \over M_1 M_2 M_3 } + {n \over g} \\ 0 
\end{array}
  \right](0,\tau M_1 M_2 M_3),
\nonumber
\end{eqnarray}
where $\ell_0$ is an integer corresponding 
to a particular solution of $M_3 l_0 = M_1 m_0 + i + j - k $ 
with integer $m_0$.

Zero-mode wavefunctions on the orbifold with the magnetic flux are 
obtained as even or odd linear combinations of 
wavefunctions on the torus with the magnetic 
flux (\ref{eq:even-odd-function}).
Thus, it is straightforward to extend the above computations of
Yukawa couplings on the torus to Yukawa couplings 
on the orbifold.
As a result, Yukawa couplings on the orbifold are obtained 
as proper linear combinations of Yukawa couplings 
on the torus, i.e. linear combinations of $\vartheta$ functions.
Here we introduce the following short notation for 
the Yukawa coupling,
\begin{eqnarray}
\eta_{N} &=& \vartheta
\left[\begin{array}{c} 
\frac{N}{M} \\ 0
\end{array} \right] \left(0,\tau M \right), 
\label{eq:notation-y}
\end{eqnarray}
where 
\begin{eqnarray}
M &=& M_1M_2M_3. \qquad 
%N \ = \ M_2i-M_1j+M_1M_2m. 
\nonumber
\end{eqnarray}
Since the value of $M$ is unique in one model, 
we omit the value of $M$ as well as $\tau$ 
for a compact presentation of long equations.
%but we will show its value in each model 
%in the following subsection and Appendix. 

Four models in Table \ref{higgs} has $|I^{(1)}_{bc}|=0$, 
where the Higgs zero-mode corresponds to the even 
function, that is, the constant profile.
We can repeat the above calculation for this case,  
that is, the case where, one of wavefunctions in (\ref{eq:yukawa}), e.g. 
$\Theta^{i,M_1}(z)$ is constant.
As a result, the Yukawa matrix is proportional to 
the $(3 \times 3)$ unit matrix, $Y_{jk} = c' \delta_{jk}$.
That is not realistic.
Thus, we will not consider such models.

At any rate, we can apply the above selection rule and $\eta_{N}$ 
for 20 classes of models, which have been classified 
in section 3, 
in order to analyze explicitly all of possible 
patterns of Yukawa matrices.
In the next subsection, we show one example of 
Yukawa matrix among 20 classes of models.
In Appendix, we show all of possible Yukawa 
matrices for 15 classes of models in Table \ref{higgs} 
except models with $I^{(1)}_{bc}=0$ and the model without zero-modes 
for the Higgs fields. 

\begin{table}[t]
\begin{center}
\begin{tabular}{c|c|c|c}
& $L_i (\lambda^{ab})$ & $R_j(\lambda^{ca})$ & $H_k(\lambda^{bc})$ 
\\ \hline
0 & 
$\frac{1}{\sqrt{2}}\left(\Theta^{1,7}-\Theta^{6,7}\right)$ & 
$\frac{1}{\sqrt{2}}\left(\Theta^{1,7}-\Theta^{6,7}\right)$ & 
$\Theta^{0,14}$ \\  
1 & 
$\frac{1}{\sqrt{2}}\left(\Theta^{2,7}-\Theta^{5,7}\right)$ & 
$\frac{1}{\sqrt{2}}\left(\Theta^{2,7}-\Theta^{5,7}\right)$ & 
$\frac{1}{\sqrt{2}}\left(\Theta^{1,14}+\Theta^{13,14}\right)$ \\  
2 & 
$\frac{1}{\sqrt{2}}\left(\Theta^{3,7}-\Theta^{4,7}\right)$ & 
$\frac{1}{\sqrt{2}}\left(\Theta^{3,7}-\Theta^{4,7}\right)$ & 
$\frac{1}{\sqrt{2}}\left(\Theta^{2,14}+\Theta^{12,14}\right)$ \\  
3 & - & - & 
$\frac{1}{\sqrt{2}}\left(\Theta^{3,14}+\Theta^{11,14}\right)$ \\  
4 & - & - & 
$\frac{1}{\sqrt{2}}\left(\Theta^{4,14}+\Theta^{10,14}\right)$ \\  
5 & - & - & 
$\frac{1}{\sqrt{2}}\left(\Theta^{5,14}+\Theta^{9,14}\right)$ \\  
6 & - & - & 
$\frac{1}{\sqrt{2}}\left(\Theta^{6,14}+\Theta^{8,14}\right)$ \\  
7 & - & - & 
$\Theta^{7,14}$ \\  
\end{tabular}
\end{center}
\caption{Zero-mode wavefunctions in the 7-7-14 model.}
\label{7-7-14model}
\end{table}

\subsection{An illustrating example: 7-7-14 model}
\label{ssec:7-7-14}

Let us study the model with 
$(|I^{(1)}_{ab}|,|I^{(1)}_{ca}|,|I^{(1)}_{bc}|)=(7,7,14)$.
Following Table \ref{higgs}, we consider the combination of zero-mode 
wavefunctions, where zero-modes of 
left and right-handed matter fields and Higgs fields correspond to 
odd, odd and even wavefunctions, respectively. 
Their wavefunctions are shown in Table~\ref{7-7-14model}.
Hereafter, for concreteness, 
we denote left and right-handed matter fields 
and Higgs fields by $L_i$, $R_j$ and $H_k$, respectively.
This model has eight zero-modes for Higgs fields.

Then, their Yukawa couplings $Y_{ijk}L_iR_jH_k$ are 
written by 
\begin{eqnarray}
Y_{ijk}H_k &=& 
  y_{ij}^0 H_0 + y_{ij}^1 H_1 + y_{ij}^2 H_2 
+ y_{ij}^3 H_3 + y_{ij}^4 H_4 + y_{ij}^5 H_5
+ y_{ij}^6 H_6 + y_{ij}^7 H_7, 
\nonumber
%\label{eq:yijkhk}
\end{eqnarray}
where 
\begin{eqnarray}
y_{ij}^0 &=& 
\left( \begin{array}{ccc}
-y_c & 0 & 0 \\
0 & -y_e & 0 \\
0 & 0 & -y_g 
\end{array} \right), \quad
y_{ij}^1 \ = \ 
\left( \begin{array}{ccc}
0 & -\frac{1}{\sqrt{2}}y_d & 0 \\
-\frac{1}{\sqrt{2}}y_d & 0 & -\frac{1}{\sqrt{2}}y_f \\
0 & -\frac{1}{\sqrt{2}}y_f & \frac{1}{\sqrt{2}}y_h  
\end{array} \right), 
\nonumber \\
y_{ij}^2 &=& 
\left( \begin{array}{ccc}
\frac{1}{\sqrt{2}}y_a & 0 &  -\frac{1}{\sqrt{2}}y_e  \\
0 & 0 & \frac{1}{\sqrt{2}}y_g \\ 
-\frac{1}{\sqrt{2}}y_e & \frac{1}{\sqrt{2}}y_g & 0   
\end{array} \right), \quad 
y_{ij}^3 \ = \ 
\left( \begin{array}{ccc}
0 & \frac{1}{\sqrt{2}}y_b  &  \frac{1}{\sqrt{2}}y_f \\
\frac{1}{\sqrt{2}}y_b & \frac{1}{\sqrt{2}}y_h & 0 \\ 
\frac{1}{\sqrt{2}}y_f & 0 & 0 
\end{array} \right), 
\nonumber \\
y_{ij}^4 &=& 
\left( \begin{array}{ccc}
0 & \frac{1}{\sqrt{2}}y_g  &  \frac{1}{\sqrt{2}}y_c \\
\frac{1}{\sqrt{2}}y_g & \frac{1}{\sqrt{2}}y_a & 0 \\ 
\frac{1}{\sqrt{2}}y_c & 0 & 0 
\end{array} \right), \quad 
y_{ij}^5 \ = \ 
\left( \begin{array}{ccc}
\frac{1}{\sqrt{2}}y_h & 0 &  -\frac{1}{\sqrt{2}}y_d  \\
0 & 0 & \frac{1}{\sqrt{2}}y_b \\ 
-\frac{1}{\sqrt{2}}y_d & \frac{1}{\sqrt{2}}y_b & 0   
\end{array} \right), 
\nonumber \\
y_{ij}^6 &=& 
\left( \begin{array}{ccc}
0 & -\frac{1}{\sqrt{2}}y_e & 0 \\
-\frac{1}{\sqrt{2}}y_e & 0 & -\frac{1}{\sqrt{2}}y_c \\
0 & -\frac{1}{\sqrt{2}}y_c & \frac{1}{\sqrt{2}}y_a  
\end{array} \right), \quad 
y_{ij}^7 \ = \ 
\left( \begin{array}{ccc}
-y_f & 0 & 0 \\
0 & -y_d & 0 \\
0 & 0 & -y_b 
\end{array} \right), 
\label{eq:yij}
\end{eqnarray}
and 
\begin{eqnarray}
y_a &=& 
\eta_0 +2\eta_{98} +2\eta_{196} +2\eta_{294}, 
\nonumber\\
y_b &=& 
\eta_{7} + \eta_{91} + \eta_{105} + \eta_{189} 
+ \eta_{203} + \eta_{287} +\eta_{301}, 
\nonumber \\ 
y_c &=& 
\eta_{14} + \eta_{84} + \eta_{112} + \eta_{182} 
+ \eta_{210} + \eta_{280} +\eta_{308},  
\nonumber \\
y_d &=& 
\eta_{21} + \eta_{77} + \eta_{119} + \eta_{175} 
+ \eta_{217} + \eta_{273} +\eta_{315}, 
\nonumber \\ 
y_e &=& 
\eta_{28} + \eta_{70} + \eta_{126} + \eta_{168} 
+ \eta_{224} + \eta_{266} +\eta_{322},  
\nonumber \\
y_f &=& 
\eta_{35} + \eta_{63} + \eta_{133} + \eta_{161} 
+ \eta_{231} + \eta_{259} +\eta_{329}, 
\nonumber \\ 
y_g &=& 
\eta_{42} + \eta_{56} + \eta_{140} + \eta_{154} 
+ \eta_{238} + \eta_{252} +\eta_{336},  
\nonumber \\
y_h &=& 
2\eta_{49} +2\eta_{147} +2\eta_{245} + \eta_{343}. 
\nonumber
\end{eqnarray}
Here we have used the short notation $\eta_N$ defined in Eq.~(\ref{eq:notation-y}) 
with the omitted value $M=M_1M_2M_3=686$.

\subsection{Numerical examples in 7-7-14 model}
\label{ssec:numerical}

\begin{figure}[t]
\begin{center}
\begin{minipage}{0.45\linewidth}
\begin{center}
\epsfig{file=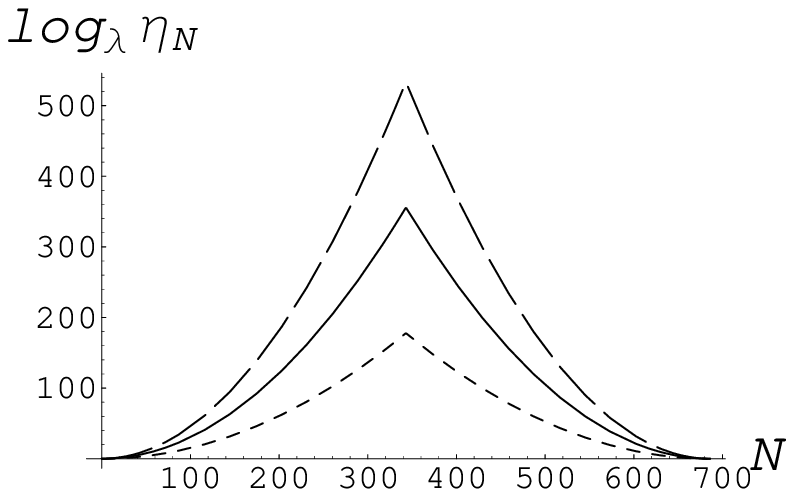,width=\linewidth}
\end{center}
\end{minipage}
\begin{minipage}{0.45\linewidth}
\begin{center}
\epsfig{file=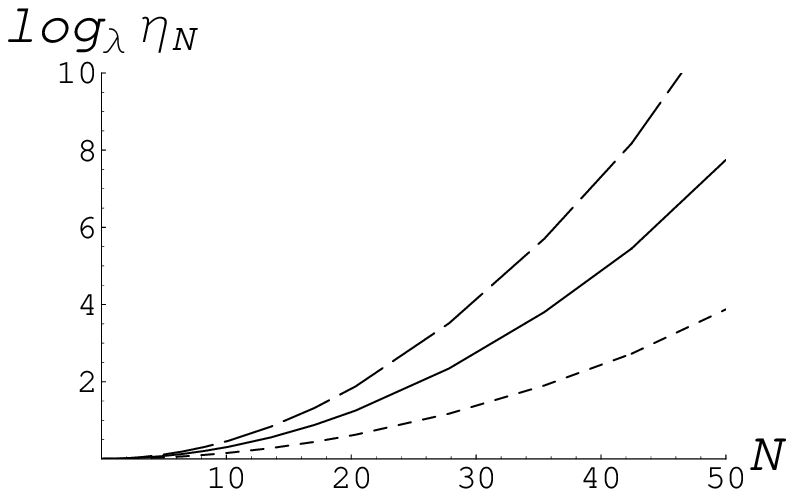,width=\linewidth}
\end{center}
\end{minipage}
\end{center}
\caption{The $N$-dependence of $\log_\lambda \eta_N$ 
in the 7-7-14 model ($M=686$), where $\lambda=0.22$ 
is chosen to the Cabibbo angle. 
The solid, dashed and dotted curves correspond to 
$\tau=i$, $1.5i$ and $0.5i$, respectively. 
Note that $\eta_N$ has a periodicity 
$\eta_{N+nM}=\eta_N$ with an integer $n$.}
\label{fig:eta}
\end{figure}

Here, we give examples of numerical studies by using the 7-7-14 model, 
which is discussed in the previous subsection.
For such studies, the numerical values of $\eta_N$ defined in 
Eq.~(\ref{eq:notation-y}) are useful. 
The $N$-dependence of $\eta_N$ is shown in Fig.~\ref{fig:eta}. 

We assume that both the up-sector and the down-sector of quarks 
as well as their Higgs fields have the Yukawa matrix, 
which is led in the 7-7-14 model.
Such situation is realized in the case that 
we start with the $U(8)$ gauge group and break  
it to $U(4) \times U(2)_L \times U(2)_R$ by the magnetic flux, 
and then the Pati-Salam gauge group is 
broken to the Standard gauge group 
by assuming VEVs of Higgs fields on fixed points.
Alternatively, we break the $U(8)$ gauge group 
to $U(3)\times U(1)_1 \times U(2)_L \times U(1)_2 \times U(1)_3$ 
by magnetic fluxes and orbifold projections 
as discussed in section 3.
Then, both the up-sector and down-sector 
of quarks can correspond to the Yukawa matrix 
led in the 7-7-14 model, although the 
up-sector and down-sector can generically 
correspond to different patterns of Yukawa matrices.
In both cases, VEVs of the up-sector and down-sector 
Higgs fields are independent.

First, we consider the case that 
VEVs of $H_d^6$, $H_d^7$ and $H_u^0$ 
are non-vanishing and the other VEVs vanish.
In this case, the relevant Yukawa couplings are 
\begin{eqnarray}
Y^u_{ijk}H_k &=& 
\left( \begin{array}{ccc}
-y_c & &  \\
& -y_e & \\
 & & -y_g 
\end{array} \right) H_u^0, 
\nonumber \\
Y^d_{ijk}H_k &=& 
\left( \begin{array}{ccc}
-y_f H_d^7& -\frac{1}{\sqrt{2}}y_e H_d^6& 0 \\
-\frac{1}{\sqrt{2}}y_e H_d^6& -y_d H_d^7& -\frac{1}{\sqrt{2}}y_c H_d^6 \\
0 & -\frac{1}{\sqrt{2}}y_c H_d^6& \frac{1}{\sqrt{2}}y_aH_d^6 -y_b H_d^7
\end{array} \right). 
\nonumber
\end{eqnarray}
Let us assume $\langle H_d^6\rangle =-\langle H_d^7 \rangle $ for their VEVs.
Then, quark mass ratios are obtained from these matrices as
\begin{eqnarray}
(m_u, m_c, m_t)/m_t &\sim& 
(7.6 \times 10^{-4},\, 6.8 \times 10^{-2},\, 1.0), 
\nonumber \\
(m_d, m_s, m_b)/m_b &\sim& 
(7.5 \times 10^{-4},\, 5.1\times 10^{-2},\, 1.0), 
\nonumber
\end{eqnarray}
for $\tau =i$.
Furthermore, the mixing angles are obtained as 
\begin{eqnarray}
|V_{CKM}| &\sim& 
\left( \begin{array}{ccc}
0.97 & 0.24 & 0.0025 \\
0.24 & 0.95 & 0.20 \\
0.046 & 0.19 & 0.98 
\end{array} \right). 
\nonumber
\end{eqnarray}
Similarly, for $\tau = 1.5 i$, 
quark mass ratios are obtained as 
\begin{eqnarray}
(m_u, m_c, m_t)/m_t &\sim& 
(2.1 \times 10^{-5},\, 1.8 \times 10^{-2},\, 1.0), 
\nonumber \\
(m_b,m_s,m_d)/m_b &\sim& 
(1.4 \times 10^{-4},\, 1.7 \times 10^{-2},\, 1.0), 
\nonumber
\end{eqnarray}
and the mixing angles are obtained as 
\begin{eqnarray}
|V_{CKM}| &\sim& 
\left( \begin{array}{ccc}
0.99 & 0.13 & 0.00029 \\
0.13 & 0.98 & 0.13 \\
0.017 & 0.13 & 0.99 
\end{array} \right). 
\nonumber
\end{eqnarray}

Let us consider another type of VEVs.
We assume that VEVs of $H_u^0$, $H_u^2$, $H_d^1$ and 
$H_d^7$ are non-vanishing and the other VEVs vanish.
Furthermore, we consider the case with 
$\langle H_u^0 \rangle = -\langle H_u^2\rangle $ 
and $\langle H_d^1 \rangle = \langle H_d^7 \rangle /3$. 
In this case, the mass ratios are given by 
\begin{eqnarray}
(m_u, m_c, m_t)/m_t &\sim& 
(2.9 \times 10^{-5},\, 2.5 \times 10^{-2},\, 1.0), 
\nonumber \\
(m_d, m_s, m_b)/m_b &\sim& 
(4.4 \times 10^{-3},\, 0.18,\, 1.0), 
\nonumber
\end{eqnarray}
for $\tau =i$, and the mixing angles are given by 
\begin{eqnarray}
|V_{CKM}| &\sim& 
\left( \begin{array}{ccc}
0.98 & 0.22 & 0.018 \\
0.22 & 0.98 & 0.0014 \\
0.017 & 0.0052  & 1.0
\end{array} \right). 
\nonumber
\end{eqnarray}
Similarly, for $\tau =1.5i$ the mass ratios 
and the mixing angles are given by
\begin{eqnarray}
(m_u, m_c, m_t)/m_t &\sim& 
(5.6\times 10^{-6},\, 4.7 \times 10^{-3},\, 1.0), 
\nonumber \\
(m_d, m_s, m_b)/m_b &\sim& 
(3.3 \times 10^{-3},\, 7.1 \times 10^{-2},\, 1.0), 
\nonumber 
\end{eqnarray}
\begin{eqnarray}
|V_{CKM}| &\sim& 
\left( \begin{array}{ccc}
0.98 & 0.22  & 0.0034 \\
0.22 & 0.98 & 0.000081 \\
0.0033 & 0.00081 & 1.0 
\end{array} \right). 
\nonumber
\end{eqnarray}
Thus, these values can realize experimental values 
of quark masses and mixing angles at a certain level 
by using a few parameters, i.e. $\tau$ and 
a couple of VEVs of Higgs fields.
If we consider more non-vanishing VEVs of Higgs fields, 
we could obtain more realistic values.
For example, we assume that VEVs of $H_u^0$, $H_u^1$, $H_u^2$, 
$H_d^1$ and $H_d^7$ are non-vanishing and they satisfy 
$-\langle H_u^0 \rangle = \langle H_u^1\rangle = \langle H_u^2\rangle $ 
and $\langle H_d^1 \rangle = - \langle H_d^7 \rangle /2$
while the other VEVs vanish. 
For $\tau = 1.5 i$, we realize the mass ratios, 
$m_u/m_t \sim 2.7 \times 10^{-5}$, $m_c/m_t \sim 3.5 \times 10^{-3}$, 
 $m_d/m_b \sim 7.3 \times 10^{-3}$ and  $m_s/m_b \sim 7.5 \times 10^{-2}$, 
and mixing angles, $V_{us} \sim 0.2$, $V_{cb} \sim 0.03$ and 
$V_{ub} \sim 0.006$. 
When we consider more non-vanishing VEVs of Higgs fields, 
it is possible to derive completely realistic values.
Similarly, we can study other classes of models 
and they have a rich flavor structure.

\section{Conclusion}

We have studied three generation magnetized orbifold models.
We have classified their flavor structures 
and studied explicitly possible patterns of Yukawa matrices.
Our models have a rich flavor structure, especially compared 
with the corresponding models without orbifolding.
Realistic quark masses and mixing angles can be 
derived within the framework of magnetized orbifold models.
We can extend our numerical studies including 
the lepton sector.

Here, we have studied the models, where all of three 
generations are originated from bulk modes.
However, we have degree of freedom to put  
some of three generations of quarks and leptons 
on certain orbifold fixed points.
In addition, we can assume that some Higgs fields 
are localized on certain orbifold fixed points.
In such cases, we would have more variety of 
flavor structure.
Furthermore, it is possible to consider 
localized magnetic fluxes on orbifold 
fixed points, which are independent of 
the bulk magnetic flux.\footnote{
See e.g. \cite{Lee:2003mc}.}
Since such localized magnetic fluxes would affect 
profiles of zero-modes, that is one 
of interesting extensions of our models.

We have restricted ourselves to Abelian fluxes, 
but we can also extend our analysis to models with 
non-Abelian fluxes, which can reduce ranks of gauge groups.
Moreover, although we have concentrated on 
the factorizable torus, $(T^2)^3$, it would be 
interesting to study possibilities for extensions to 
non-factorizable orbifolds \cite{Forste:2006wq}.

\subsection*{Acknowledgement}
H.~A. is supported by the Grant-in-Aid for the Global COE Program 
``Weaving Science Web beyond Particle-matter Hierarchy'' from the 
Ministry of Education, Culture, Sports, Science and Technology of Japan. 
K.-S.~C. and T.~K. are supported in part by the Grant-in-Aid for 
Scientific Research No.~20$\cdot$08326 and No.~20540266 from the 
Ministry of Education, Culture, Sports, Science and Technology of Japan.
T.~K. is also supported in part by the Grant-in-Aid for the Global COE 
Program "The Next Generation of Physics, Spun from Universality and 
Emergence" from the Ministry of Education, Culture,Sports, Science and 
Technology of Japan.

\appendix

\section{Possible patterns of Yukawa matrices}

In this appendix, we show explicitly all of possible Yukawa 
matrices for 15 classes of models in Table \ref{higgs} 
except the models with $I^{(1)}_{bc}=0$ and the model without zero-modes 
for the Higgs fields.

\subsection{(Even-Even-Even) wavefunctions}

Here, we study the patterns of Yukawa matrices 
in the models, where zero-modes of left, right-handed 
matter fields and Higgs fields correspond to 
even, even and even functions, respectively.

\subsubsection{4-4-8 model}

Let us study the model with 
$(|I^{(1)}_{ab}|,|I^{(1)}_{ca}|,|I^{(1)}_{bc}|)=(4,4,8)$.
The following table shows zero-mode wavefunctions of left, right-handed 
matter fields and Higgs fields.
\begin{center}
\begin{tabular}{c|c|c|c}
& $L_i (\lambda^{ab})$ & $R_j (\lambda^{ca})$ & $H_k (\lambda^{bc})$ 
\\ \hline
0 & 
$\Theta^{0,4}$ & $\Theta^{0,4}$ & $\Theta^{0,8}$ \\  
1 & 
$\frac{1}{\sqrt{2}}\left(\Theta^{1,4}+\Theta^{3,4}\right)$ & 
$\frac{1}{\sqrt{2}}\left(\Theta^{1,4}+\Theta^{3,4}\right)$ & 
$\frac{1}{\sqrt{2}}\left(\Theta^{1,8}+\Theta^{7,8}\right)$ \\  
2 & 
$\Theta^{2,4}$ & 
$\Theta^{2,4}$ & 
$\frac{1}{\sqrt{2}}\left(\Theta^{2,8}+\Theta^{6,8}\right)$ \\  
3 & - & - & 
$\frac{1}{\sqrt{2}}\left(\Theta^{3,8}+\Theta^{5,8}\right)$ \\  
4 & - & - & 
$\Theta^{4,8}$ \\  
\end{tabular}
\end{center}
This model has five zero-modes for the Higgs fields.
Yukawa couplings $Y_{ijk} L_i R_j H_k$ are given by 
\begin{eqnarray}
Y_{ijk}H_k &=& 
\left( \begin{array}{ccc}
y_a H_0 + y_e H_4 & y_4 H_3 + y_b H_1 & y_c H_2 \\
y_4 H_3 + y_b H_1 & \frac{1}{\sqrt{2}}(y_a+y_e)H_2 
+ y_c(H_0+H_4) & y_b H_3+y_d H_1 \\
y_c H_2 & y_b H_3+y_d H_1 & y_e H_0 + y_a H_4 
\end{array} \right), 
\nonumber
\end{eqnarray}
where
\begin{eqnarray}
\begin{array}{rclcrcl}
y_a &=& \eta_0 + 2\eta_{32}+ \eta_{64}, & & 
y_b &=& \eta_{4} + \eta_{28} + \eta_{36}+ \eta_{60}, 
\nonumber \\
y_c &=& \eta_{8} + \eta_{24} + \eta_{40}+ \eta_{56}, & & 
y_d &=& \eta_{12} + \eta_{20} + \eta_{44} + \eta_{52}, 
\nonumber \\
y_e &=& 2\eta_{16} + 2\eta_{48}, & & & & 
\end{array}
\nonumber
\end{eqnarray}
in the short notation $\eta_N$ defined in Eq.~(\ref{eq:notation-y}) with 
$M=M_1M_2M_3=128$.

\subsubsection{4-5-9 model}

Here we show the model with 
$(|I^{(1)}_{ab}|,|I^{(1)}_{ca}|,|I^{(1)}_{bc}|)=(4,5,9)$.
The following table shows zero-mode wavefunctions of left, right-handed 
matter fields and Higgs fields.
\begin{center}
\begin{tabular}{c|c|c|c}
& $L_i (\lambda^{ab})$ & $R_j (\lambda^{ca})$ & $H_k (\lambda^{bc})$ 
\\ \hline
0 & 
$\Theta^{0,4}$ & $\Theta^{0,5}$ & $\Theta^{0,9}$ \\  
1 & 
$\frac{1}{\sqrt{2}}\left(\Theta^{1,4}+\Theta^{3,4}\right)$ & 
$\frac{1}{\sqrt{2}}\left(\Theta^{1,5}+\Theta^{4,5}\right)$ & 
$\frac{1}{\sqrt{2}}\left(\Theta^{1,9}+\Theta^{8,9}\right)$ \\  
2 & 
$\Theta^{2,4}$ & 
$\frac{1}{\sqrt{2}}\left(\Theta^{2,5}+\Theta^{3,5}\right)$ & 
$\frac{1}{\sqrt{2}}\left(\Theta^{2,9}+\Theta^{7,9}\right)$ \\  
3 & - & - & 
$\frac{1}{\sqrt{2}}\left(\Theta^{3,9}+\Theta^{6,9}\right)$ \\  
4 & - & - & 
$\frac{1}{\sqrt{2}}\left(\Theta^{4,9}+\Theta^{5,9}\right)$ \\  
\end{tabular}
\end{center}
This model has five zero-modes for Higgs fields.
Yukawa couplings $Y_{ijk} L_i R_j H_k$ are given by 
\begin{eqnarray}
Y_{ijk}H_k &=& 
y_{ij}^0 H_0 + y_{ij}^1 H_1 + y_{ij}^2 H_2 
+ y_{ij}^3 H_3 + y_{ij}^4 H_4, 
\nonumber
\end{eqnarray}
where
\begin{eqnarray}
y_{ij}^0 &=& 
\left( \begin{array}{ccc}
\eta_0 & \sqrt{2}\eta_{36} & \sqrt{2}\eta_{72} \\
\sqrt{2}\eta_{45} & \eta_{9}+\eta_{81} & \eta_{27}+\eta_{63} \\
\eta_{90} & \sqrt{2}\eta_{54} & \sqrt{2}\eta_{18}
\end{array} \right), 
\nonumber \\ 
y_{ij}^1 &=& 
\left( \begin{array}{ccc}
\frac{1}{\sqrt{2}}(\eta_{20}+\eta_{40}) & \eta_{4}+\eta_{76} & \eta_{32}+\eta_{68} \\
\eta_{5}+\eta_{85} & \frac{1}{\sqrt{2}}(\eta_{31}+\eta_{41}+\eta_{49}+\eta_{59}) 
& \frac{1}{\sqrt{2}}(\eta_{13}+\eta_{23}+\eta_{67}+\eta_{77}) \\
\sqrt{2}\eta_{50} & \eta_{44}+\eta_{64} & \eta_{22}+\eta_{58} 
\end{array} \right), 
\nonumber \\
y_{ij}^2 &=& 
\left( \begin{array}{ccc}
\frac{1}{\sqrt{2}}(\eta_{20}+\eta_{40}) & \eta_{44}+\eta_{64} & \eta_{8}+\eta_{28} \\
\eta_{35}+\eta_{55} & \frac{1}{\sqrt{2}}(\eta_{1}+\eta_{19}+\eta_{71}+\eta_{89}) 
& \frac{1}{\sqrt{2}}(\eta_{17}+\eta_{37}+\eta_{53}+\eta_{73}) \\
\sqrt{2}\eta_{10} & \eta_{26}+\eta_{46} & \eta_{62}+\eta_{82}
\end{array} \right), 
\nonumber \\ 
y_{ij}^3 &=& 
\left( \begin{array}{ccc}
\frac{1}{\sqrt{2}}(\eta_{60}+\eta_{80}) & \eta_{24}+\eta_{84} & \eta_{12}+\eta_{48} \\
\eta_{15}+\eta_{75} & \frac{1}{\sqrt{2}}(\eta_{21}+\eta_{39}+\eta_{51}+\eta_{69}) 
& \frac{1}{\sqrt{2}}(\eta_{3}+\eta_{33}+\eta_{57}+\eta_{87}) \\
\sqrt{2}\eta_{30} & \eta_{6}+\eta_{26} & \eta_{42}+\eta_{78}
\end{array} \right), 
\nonumber \\
y_{ij}^4 &=& 
\left( \begin{array}{ccc}
\frac{1}{\sqrt{2}}(\eta_{60}+\eta_{80}) & \eta_{16}+\eta_{56} & \eta_{52}+\eta_{88} \\
\eta_{25}+\eta_{65} & \frac{1}{\sqrt{2}}(\eta_{11}+\eta_{29}+\eta_{61}+\eta_{79}) 
& \frac{1}{\sqrt{2}}(\eta_{7}+\eta_{43}+\eta_{47}+\eta_{83}) \\
\sqrt{2}\eta_{70} & \eta_{34}+\eta_{74} & \eta_{2}+\eta_{38} 
\end{array} \right), 
\nonumber 
\end{eqnarray}
in the short notation $\eta_N$ defined in Eq.~(\ref{eq:notation-y}) with 
$M=M_1M_2M_3=180$.

\subsubsection{4-5-1 model}

Here we show the model with 
$(|I^{(1)}_{ab}|,|I^{(1)}_{ca}|,|I^{(1)}_{bc}|)=(4,5,1)$.
The following table shows zero-mode wavefunctions of left, right-handed 
matter fields and Higgs field.
\begin{center}
\begin{tabular}{c|c|c|c}
& $L_i (\lambda^{ab})$ & $R_j (\lambda^{ca})$ & $H_k (\lambda^{bc})$ 
\\ \hline
0 & $\Theta^{0,4}$ & $\Theta^{0,5}$ & $\Theta^{0,1}$ \\  
1 & $\frac{1}{\sqrt{2}}\left(\Theta^{1,4}+\Theta^{3,4}\right)$ & 
$\frac{1}{\sqrt{2}}\left(\Theta^{1,5}+\Theta^{4,5}\right)$ &  \\  
2 & $\Theta^{2,4}$ & 
$\frac{1}{\sqrt{2}}\left(\Theta^{2,5}+\Theta^{3,5}\right)$ &  \\  
\end{tabular}
\end{center}
This model has a single zero-modes for the Higgs field.
Yukawa couplings $Y_{ijk} L_i R_j H_k$ are given 
\begin{eqnarray}
Y_{ijk}H_k &=& 
\left( \begin{array}{ccc}
y_0 & \sqrt{2}\eta_4 & \sqrt{2}\eta_8 \\
\sqrt{2}\eta_5 & (\eta_1+\eta_9) & (\eta_3+\eta_7) \\
\eta_{10} & \sqrt{2}\eta_6 & \sqrt{2}\eta_2 
\end{array} \right)H_0.
\nonumber
\end{eqnarray}
Here we have used the short notation $\eta_N$ defined in Eq.~(\ref{eq:notation-y}) 
with the omitted value $M=M_1M_2M_3=20$.

\subsubsection{5-5-10 model}

Here we show the model with 
$(|I^{(1)}_{ab}|,|I^{(1)}_{ca}|,|I^{(1)}_{bc}|)=(5,5,10)$.
The following table shows zero-mode wavefunctions of left, right-handed 
matter fields and Higgs fields.
\begin{center}
\begin{tabular}{c|c|c|c}
& $L_i (\lambda^{ab})$ & $R_j (\lambda^{ca})$ & $H_k (\lambda^{bc})$ 
\\ \hline
0 & 
$\Theta^{0,5}$ & $\Theta^{0,5}$ & $\Theta^{0,10}$ \\  
1 & 
$\frac{1}{\sqrt{2}}\left(\Theta^{1,5}+\Theta^{4,5}\right)$ & 
$\frac{1}{\sqrt{2}}\left(\Theta^{1,5}+\Theta^{4,5}\right)$ & 
$\frac{1}{\sqrt{2}}\left(\Theta^{1,10}+\Theta^{9,10}\right)$ \\  
2 & 
$\frac{1}{\sqrt{2}}\left(\Theta^{2,5}+\Theta^{3,5}\right)$ & 
$\frac{1}{\sqrt{2}}\left(\Theta^{2,5}+\Theta^{3,5}\right)$ & 
$\frac{1}{\sqrt{2}}\left(\Theta^{2,10}+\Theta^{8,10}\right)$ \\  
3 & - & - & 
$\frac{1}{\sqrt{2}}\left(\Theta^{3,10}+\Theta^{7,10}\right)$ \\  
4 & - & - & 
$\frac{1}{\sqrt{2}}\left(\Theta^{4,10}+\Theta^{6,10}\right)$ \\  
5 & - & - & 
$\Theta^{5,10}$ \\  
\end{tabular}
\end{center}
This model has six zero-modes for Higgs fields.
Yukawa couplings $Y_{ijk} L_i R_j H_k$ are obtained as
\begin{eqnarray}
\lefteqn{Y_{ijk}H_k} 
\nonumber \\ &=& 
\left( \begin{array}{ccc}
y_a H_0 + y_e H_5 & y_b H_1 + y_e H_4 & y_c H_2 + y_d H_3 \\
y_b H_1 + y_e H_4 & y_c H_0 + \frac{1}{\sqrt{2}}(y_a H_2+y_f H_3) + y_d H_5 
& \frac{1}{\sqrt{2}}(y_d H_1+y_e H_2 + y_b H_3 + y_c H_4) \\
y_c H_2 + y_d H_3 & \frac{1}{\sqrt{2}}(y_d H_1+y_e H_2 + y_b H_3 + y_c H_4)
& y_b H_0 + \frac{1}{\sqrt{2}}(y_f H_1 + y_a H_4) + y_a H_5 
\end{array} \right). 
\nonumber
\end{eqnarray}
\begin{eqnarray}
\begin{array}{rclcrcl}
y_a &=& \eta_0 + 2\eta_{50}+ 2\eta_{100}, & & 
y_b &=& \eta_{5} + \eta_{45} + \eta_{55} + \eta_{95} + \eta_{105}, 
\nonumber \\
y_c &=& \eta_{10} + \eta_{40} + \eta_{60} + \eta_{90} + \eta_{110}, & & 
y_d &=& \eta_{15} + \eta_{35} + \eta_{65} + \eta_{85} + \eta_{115}, 
\nonumber \\ 
y_e &=& \eta_{20} + \eta_{30} + \eta_{70} + \eta_{80} + \eta_{120}, & & 
y_f &=& 2\eta_{25} + 2\eta_{75} + \eta_{125}, 
\end{array}
\nonumber 
\end{eqnarray}
in the short notation $\eta_N$ defined in Eq.~(\ref{eq:notation-y}) with 
$M=M_1M_2M_3=250$.

\subsection{(Even-Odd-Odd) wavefunctions}

Here, we study the patterns of Yukawa matrices 
in the models, where zero-modes of left, right-handed 
matter fields and Higgs fields correspond to 
even, odd and odd functions, respectively.

\subsubsection{4-7-11 model}

Here we show the model with 
$(|I^{(1)}_{ab}|,|I^{(1)}_{ca}|,|I^{(1)}_{bc}|)=(4,7,11)$.
The following table shows zero-mode wavefunctions of left, right-handed 
matter fields and Higgs fields.
\begin{center}
\begin{tabular}{c|c|c|c}
& $L_i (\lambda^{ab})$ & $R_j (\lambda^{ca})$ & $H_k (\lambda^{bc})$ 
\\ \hline
0 & 
$\Theta^{0,4}$ & $\frac{1}{\sqrt{2}}(\Theta^{1,7}-\Theta^{6,7})$ & 
$\frac{1}{\sqrt{2}}(\Theta^{1,11}-\Theta^{10,11})$ \\  
1 & 
$\frac{1}{\sqrt{2}}\left(\Theta^{1,4}+\Theta^{3,4}\right)$ & 
$\frac{1}{\sqrt{2}}\left(\Theta^{2,7}-\Theta^{5,7}\right)$ & 
$\frac{1}{\sqrt{2}}\left(\Theta^{2,11}-\Theta^{9,11}\right)$ \\  
2 & 
$\Theta^{2,4}$ & 
$\frac{1}{\sqrt{2}}\left(\Theta^{3,7}-\Theta^{4,7}\right)$ & 
$\frac{1}{\sqrt{2}}\left(\Theta^{3,11}-\Theta^{8,11}\right)$ \\  
3 & - & - & 
$\frac{1}{\sqrt{2}}\left(\Theta^{4,11}-\Theta^{7,11}\right)$ \\  
4 & - & - & 
$\frac{1}{\sqrt{2}}\left(\Theta^{5,11}-\Theta^{6,11}\right)$ \\  
\end{tabular}
\end{center}
This model has five zero-modes for the Higgs fields.
Yukawa couplings $Y_{ijk} L_i R_j H_k$ are given by
\begin{eqnarray}
Y_{ij}^kH_k=
  y_{ij}^0 H_0 + y_{ij}^1 H_1 + y_{ij}^2 H_2 
+ y_{ij}^3 H_3 + y_{ij}^4 H_4, 
\nonumber
\end{eqnarray}
where 
\begin{eqnarray}
y_{ij}^0 &=& \frac{1}{\sqrt{2}} 
\left( \begin{array}{ccc}
\sqrt{2}(\eta_4-\eta_{136}) & 
\sqrt{2}(\eta_{92}-\eta_{48}) & 
\sqrt{2}(\eta_{128}-\eta_{40}) \\
\eta_{81}-\eta_{59}-\eta_{95}+\eta_{73} & 
\eta_{139}-\eta_{29}-\eta_{125}+\eta_{15} & 
\eta_{51}-\eta_{117}-\eta_{37}+\eta_{103} \\ 
\sqrt{2}(\eta_{150}-\eta_{18}) & 
\sqrt{2}(\eta_{62}-\eta_{16}) & 
\sqrt{2}(\eta_{26}-\eta_{114}) 
\end{array} \right), 
\nonumber \\ 
y_{ij}^1 &=& \frac{1}{\sqrt{2}} 
\left( \begin{array}{ccc}
\sqrt{2}(\eta_{80}-\eta_{52}) & 
\sqrt{2}(\eta_{8}-\eta_{36}) & 
\sqrt{2}(\eta_{96}-\eta_{124}) \\
\eta_{3}-\eta_{25}-\eta_{129}+\eta_{151} & 
\eta_{85}-\eta_{113}-\eta_{41}+\eta_{69} & 
\eta_{135}-\eta_{107}-\eta_{47}+\eta_{19} \\ 
\sqrt{2}(\eta_{74}-\eta_{102}) & 
\sqrt{2}(\eta_{146}-\eta_{118}) & 
\sqrt{2}(\eta_{58}-\eta_{30}) 
\end{array} \right), 
\nonumber \\
y_{ij}^2 &=& \frac{1}{\sqrt{2}} 
\left( \begin{array}{ccc}
\sqrt{2}(\eta_{144}-\eta_{32}) & 
\sqrt{2}(\eta_{76}-\eta_{120}) & 
\sqrt{2}(\eta_{12}-\eta_{100}) \\
\eta_{87}-\eta_{109}-\eta_{45}+\eta_{67} & 
\eta_{1}-\eta_{111}-\eta_{43}+\eta_{153} & 
\eta_{89}-\eta_{23}-\eta_{131}+\eta_{65} \\ 
\sqrt{2}(\eta_{10}-\eta_{122}) & 
\sqrt{2}(\eta_{78}-\eta_{34}) & 
\sqrt{2}(\eta_{142}-\eta_{54}) 
\end{array} \right), 
\nonumber \\
y_{ij}^3 &=& \frac{1}{\sqrt{2}} 
\left( \begin{array}{ccc}
\sqrt{2}(\eta_{148}-\eta_{104}) & 
\sqrt{2}(\eta_{148}-\eta_{104}) & 
\sqrt{2}(\eta_{72}-\eta_{16}) \\
\eta_{171}-\eta_{115}-\eta_{39}+\eta_{17} & 
\eta_{83}-\eta_{27}-\eta_{127}+\eta_{71} & 
\eta_{5}-\eta_{61}-\eta_{13}+\eta_{149} \\ 
\sqrt{2}(\eta_{94}-\eta_{38}) & 
\sqrt{2}(\eta_{6}-\eta_{50}) & 
\sqrt{2}(\eta_{82}-\eta_{138}) 
\end{array} \right), 
\nonumber \\
y_{ij}^4 &=& \frac{1}{\sqrt{2}} 
\left( \begin{array}{ccc}
\sqrt{2}(\eta_{24}-\eta_{108}) & 
\sqrt{2}(\eta_{64}-\eta_{20}) & 
\sqrt{2}(\eta_{152}-\eta_{68}) \\
\eta_{53}-\eta_{31}-\eta_{123}+\eta_{101} & 
\eta_{141}-\eta_{57}-\eta_{7}+\eta_{13} & 
\eta_{79}-\eta_{145}-\eta_{9}+\eta_{75} \\ 
\sqrt{2}(\eta_{130}-\eta_{46}) & 
\sqrt{2}(\eta_{90}-\eta_{134}) & 
\sqrt{2}(\eta_{2}-\eta_{86}) 
\end{array} \right), 
\nonumber 
\end{eqnarray}
in the short notation $\eta_N$ defined in Eq.~(\ref{eq:notation-y}) with 
$M=M_1M_2M_3=308$.

\subsubsection{4-7-3 model}

Here we show the model with 
$(|I^{(1)}_{ab}|,|I^{(1)}_{ca}|,|I^{(1)}_{bc}|)=(4,7,3)$.
The following table shows zero-mode wavefunctions of left, right-handed 
matter fields and Higgs fields.
\begin{center}
\begin{tabular}{c|c|c|c}
& $L_i (\lambda^{ab})$ & $R_j (\lambda^{ca})$ & $H_k (\lambda^{bc})$ 
\\ \hline
0 & 
$\Theta^{0,4}$ & $\frac{1}{\sqrt{2}}(\Theta^{1,7}-\Theta^{6,7})$ 
& $\frac{1}{\sqrt{2}}(\Theta^{1,3}-\Theta^{2,3})$ \\  
1 & 
$\frac{1}{\sqrt{2}}\left(\Theta^{1,4}+\Theta^{3,4}\right)$ & 
$\frac{1}{\sqrt{2}}\left(\Theta^{2,5}-\Theta^{5,7}\right)$ & 
- \\  
2 & 
$\Theta^{2,4}$ & 
$\frac{1}{\sqrt{2}}\left(\Theta^{3,5}-\Theta^{4,5}\right)$ & 
- \\  
\end{tabular}
\end{center}
This model has a single zero-modes for Higgs fields.
Yukawa couplings $Y_{ijk} L_i R_j H_k$ are obtained as 
\begin{eqnarray}
Y_{ij}^k H_k &=& \frac{1}{\sqrt{2}} H_0 
\left( \begin{array}{ccc}
\sqrt{2}(\eta_4-\eta_{32}) & 
\sqrt{2}(\eta_{20}-\eta_8) & 
\sqrt{2}(\eta_{40}-\eta_{16}) \\
\eta_{17} +\eta_{25}-\eta_{11}-\eta_{31} & 
\eta_{1} +\eta_{41}-\eta_{13}-\eta_{29} & 
\eta_{19} +\eta_{23}-\eta_{5}-\eta_{37} \\
\sqrt{2}(\eta_{38}-\eta_{10}) & 
\sqrt{2}(\eta_{22}-\eta_{34}) & 
\sqrt{2}(\eta_{2}-\eta_{26}) 
\end{array} \right), 
\nonumber   
\end{eqnarray}
in the short notation $\eta_N$ defined in Eq.~(\ref{eq:notation-y}) with 
$M=M_1M_2M_3=84$.

\subsubsection{4-8-12 model}

Here we show the model with 
$(|I^{(1)}_{ab}|,|I^{(1)}_{ca}|,|I^{(1)}_{bc}|)=(4,8,12)$.
The following table shows zero-mode wavefunctions of left, right-handed 
matter fields and Higgs fields.
\begin{center}
\begin{tabular}{c|c|c|c}
& $L_i (\lambda^{ab})$ & $R_j (\lambda^{ca})$ & $H_k (\lambda^{bc})$ \\ \hline
0 & 
$\Theta^{0,4}$ & $\frac{1}{\sqrt{2}}(\Theta^{1,8}-\Theta^{7,8})$ & 
$\frac{1}{\sqrt{2}}(\Theta^{1,12}-\Theta^{11,12})$ \\  
1 & 
$\frac{1}{\sqrt{2}}\left(\Theta^{1,4}+\Theta^{3,4}\right)$ & 
$\frac{1}{\sqrt{2}}\left(\Theta^{2,8}-\Theta^{6,8}\right)$ & 
$\frac{1}{\sqrt{2}}\left(\Theta^{2,12}-\Theta^{10,12}\right)$ \\  
2 & 
$\Theta^{2,4}$ & 
$\frac{1}{\sqrt{2}}\left(\Theta^{3,8}-\Theta^{5,8}\right)$ & 
$\frac{1}{\sqrt{2}}\left(\Theta^{3,12}-\Theta^{9,12}\right)$ \\  
3 & - & - & 
$\frac{1}{\sqrt{2}}\left(\Theta^{4,12}-\Theta^{8,12}\right)$ \\  
4 & - & - & 
$\frac{1}{\sqrt{2}}\left(\Theta^{5,12}-\Theta^{7,12}\right)$ \\  
\end{tabular}
\end{center}
This model has five zero-modes for the Higgs fields.
Yukawa couplings $Y_{ijk} L_i R_j H_k$ are given by 
\begin{eqnarray}
Y_{ij}^k H_k &=& 
  y_{ij}^0 H_0 + y_{ij}^1 H_1 + y_{ij}^2 H_2 
+ y_{ij}^3 H_3 + y_{ij}^4 H_4, 
\nonumber
\end{eqnarray}
where 
\begin{eqnarray}
y_{ij}^0 &=& 
\left( \begin{array}{ccc}
y_b & 0 & -y_{l} \\
0 & \frac{1}{\sqrt{2}}(y_{e}-y_{i}) & 0 \\ 
-y_{f} & 0 & y_{h} 
\end{array} \right), \quad
y_{ij}^1 \ = \ 
\left( \begin{array}{ccc}
0 & y_c-y_{k} & 0 \\
\frac{1}{\sqrt{2}}(y_{b}-y_{h}) & 0 & \frac{1}{\sqrt{2}}(y_{f}-y_{l}) \\ 
0 & 0 & 0
\end{array} \right), 
\nonumber \\
y_{ij}^2 &=& 
\left( \begin{array}{ccc}
-y_{j} & 0 & y_{d} \\
0 & \frac{1}{\sqrt{2}}(y_{a}-y_{m}) & 0 \\ 
y_{d} & 0 & -y_{j} 
\end{array} \right), \quad 
y_{ij}^3 \ = \ 
\left( \begin{array}{ccc}
0 & 0 & 0 \\
\frac{1}{\sqrt{2}}(y_{f}-y_{l}) & 0 & \frac{1}{\sqrt{2}}(y_{b}-y_{h}) \\ 
0 & y_c-y_{k} & 0 
\end{array} \right), 
\nonumber \\
y_{ij}^4 &=& 
\left( \begin{array}{ccc}
y_h & 0 & -y_{f} \\
0 & \frac{1}{\sqrt{2}}(y_{e}-y_{i}) & 0 \\ 
-y_{l} & 0 & y_{b} 
\end{array} \right), 
\nonumber 
\end{eqnarray}
and 
\begin{eqnarray}
\begin{array}{rclcrcl}
y_a &=& \eta_0 + \eta_{96} + \eta_{192}  + \eta_{96}, & & 
y_b &=& \eta_{4} + \eta_{100} + \eta_{188} + \eta_{92}, 
\nonumber \\ 
y_c &=& \eta_{8} + \eta_{104} + \eta_{184} + \eta_{88}, & & 
y_d &=& \eta_{12} + \eta_{108} + \eta_{180} + \eta_{84}, 
\nonumber \\ 
y_e &=& \eta_{16} + \eta_{112} + \eta_{176} + \eta_{80}, & & 
y_f &=& \eta_{20} + \eta_{116} + \eta_{172} + \eta_{76}, 
\nonumber \\ 
y_g &=& \eta_{24} + \eta_{120} + \eta_{168} + \eta_{72}, & & 
y_h &=& \eta_{28} + \eta_{124} + \eta_{164} + \eta_{68}, 
\nonumber \\ 
y_i &=& \eta_{32} + \eta_{128} + \eta_{160} + \eta_{64}, & & 
y_j &=& \eta_{36} + \eta_{132} + \eta_{156} + \eta_{60}, 
\nonumber \\ 
y_k &=& \eta_{40} + \eta_{136} + \eta_{152} + \eta_{56}, & & 
y_l &=& \eta_{44} + \eta_{140} + \eta_{148} + \eta_{52}, 
\nonumber \\ 
y_m &=& \eta_{48} + \eta_{144} + \eta_{144} + \eta_{48}, & & & & 
\end{array}
\nonumber
\end{eqnarray}
in the short notation $\eta_N$ defined in Eq.~(\ref{eq:notation-y}) with 
$M=M_1M_2M_3=384$.

\subsubsection{4-8-4 model}

Here we show the model with 
$(|I^{(1)}_{ab}|,|I^{(1)}_{ca}|,|I^{(1)}_{bc}|)=(4,8,4)$.
The following table shows zero-mode wavefunctions of left, right-handed 
matter fields and Higgs fields.
\begin{center}
\begin{tabular}{c|c|c|c}
& $L_i (\lambda^{ab})$ & $R_j (\lambda^{ca})$ & $H_k (\lambda^{bc})$ 
\\ \hline
0 & 
$\Theta^{0,4}$ & $\frac{1}{\sqrt{2}}(\Theta^{1,8}-\Theta^{7,8})$ 
& $\frac{1}{\sqrt{2}}(\Theta^{1,4}-\Theta^{3,4})$ \\  
1 & 
$\frac{1}{\sqrt{2}}\left(\Theta^{1,4}+\Theta^{3,4}\right)$ & 
$\frac{1}{\sqrt{2}}\left(\Theta^{2,7}-\Theta^{6,7}\right)$ & 
- \\  
2 & 
$\Theta^{2,4}$ & 
$\frac{1}{\sqrt{2}}\left(\Theta^{3,7}-\Theta^{5,7}\right)$ & 
- \\  
\end{tabular}
\end{center}
This model has a single zero-modes for Higgs fields.
Yukawa couplings $Y_{ijk} L_i R_j H_k$ are obtained as
\begin{eqnarray}
Y_{ij}^k H_k &=& H_0
\left( \begin{array}{ccc}
y_b  & 0 & -y_c \\
0    & \frac{1}{\sqrt{2}}(y_a - y_d) & 0 \\
-y_c & 0 & y_b 
\end{array} \right), 
\nonumber   
\end{eqnarray}
where
\begin{eqnarray}
\begin{array}{rclcrcl}
y_a &=& \eta_0 + 2\eta_{32} + \eta_{64}, & & 
y_b &=& \eta_{4} + \eta_{28} + \eta_{36} + \eta_{60}, 
\nonumber \\ 
y_c &=& \eta_{12} + \eta_{20} + \eta_{44} + \eta_{52}, & & 
y_d &=& 2\eta_{16} + 2\eta_{48}, 
\end{array}
\nonumber  
\end{eqnarray}
in the short notation $\eta_N$ defined in Eq.~(\ref{eq:notation-y}) with 
$M=M_1M_2M_3=128$.

\subsubsection{5-7-12 model}

Here we show the model with 
$(|I^{(1)}_{ab}|,|I^{(1)}_{ca}|,|I^{(1)}_{bc}|)=(5,7,12)$.
The following table shows zero-mode wavefunctions of left, right-handed 
matter fields and Higgs fields.
\begin{center}
\begin{tabular}{c|c|c|c}
& $L_i (\lambda^{ab})$ & $R_j (\lambda^{ca})$ & $H_k (\lambda^{bc})$ 
\\ \hline
0 & 
$\Theta^{0,5}$ & $\frac{1}{\sqrt{2}}(\Theta^{1,7}-\Theta^{6,7})$ & 
$\frac{1}{\sqrt{2}}(\Theta^{1,12}-\Theta^{11,12})$ \\  
1 & 
$\frac{1}{\sqrt{2}}\left(\Theta^{1,5}+\Theta^{4,5}\right)$ & 
$\frac{1}{\sqrt{2}}\left(\Theta^{2,7}-\Theta^{5,7}\right)$ & 
$\frac{1}{\sqrt{2}}\left(\Theta^{2,12}-\Theta^{10,12}\right)$ \\  
2 & 
$\frac{1}{\sqrt{2}}\left(\Theta^{2,5}+\Theta^{3,5}\right)$ &
$\frac{1}{\sqrt{2}}\left(\Theta^{3,7}-\Theta^{4,7}\right)$ & 
$\frac{1}{\sqrt{2}}\left(\Theta^{3,12}-\Theta^{9,12}\right)$ \\  
3 & - & - & 
$\frac{1}{\sqrt{2}}\left(\Theta^{4,12}-\Theta^{8,12}\right)$ \\  
4 & - & - & 
$\frac{1}{\sqrt{2}}\left(\Theta^{5,12}-\Theta^{7,12}\right)$ \\  
\end{tabular}
\end{center}
This model has five zero-modes for the Higgs fields.
Yukawa coupling $Y_{ijk}L_iR_j H_k$ are given by 
\begin{eqnarray}
Y_{ijk}H_k &=& 
  y_{ij}^0 H_0 + y_{ij}^1 H_1 + y_{ij}^2 H_2 
+ y_{ij}^3 H_3 + y_{ij}^4 H_4, 
\nonumber
\end{eqnarray}
where
\begin{eqnarray}
y_{ij}^0 &=& \frac{1}{\sqrt{2}} 
\left( \begin{array}{ccc}
\sqrt{2}(\eta_5-\eta_{65}) & 
\sqrt{2}(\eta_{185}-\eta_{115}) & 
\sqrt{2}(\eta_{55}+\eta_{125}) \\
\eta_{173}-\eta_{103}-\eta_{187}+\eta_{163} & 
\eta_{67}-\eta_{137}-\eta_{53}+\eta_{17} & 
\eta_{113}-\eta_{43}-\eta_{127}+\eta_{197} \\ 
\eta_{79}-\eta_{149}-\eta_{19}+\eta_{89} & 
\eta_{101}-\eta_{31}-\eta_{199}+\eta_{151} & 
\eta_{139}-\eta_{209}-\eta_{41}+\eta_{29} 
\end{array} \right), 
\nonumber \\ 
y_{ij}^1 &=& \frac{1}{\sqrt{2}} 
\left( \begin{array}{ccc}
\sqrt{2}(\eta_{170}-\eta_{110}) & 
\sqrt{2}(\eta_{10}-\eta_{130}) & 
\sqrt{2}(\eta_{190}+\eta_{50}) \\
\eta_{2}-\eta_{142}-\eta_{58}+\eta_{82} & 
\eta_{178}-\eta_{38}-\eta_{122}+\eta_{158} & 
\eta_{62}-\eta_{202}-\eta_{118}+\eta_{22} \\ 
\eta_{166}-\eta_{26}-\eta_{194}+\eta_{94} & 
\eta_{74}-\eta_{206}-\eta_{46}+\eta_{94} & 
\eta_{106}-\eta_{34}-\eta_{134}+\eta_{146} 
\end{array} \right), 
\nonumber \\
y_{ij}^2 &=& \frac{1}{\sqrt{2}} 
\left( \begin{array}{ccc}
\sqrt{2}(\eta_{75}-\eta_{135}) & 
\sqrt{2}(\eta_{165}-\eta_{45}) & 
\sqrt{2}(\eta_{15}-\eta_{195}) \\
\eta_{177}-\eta_{33}-\eta_{117}+\eta_{93} & 
\eta_{3}-\eta_{207}-\eta_{123}+\eta_{87} & 
\eta_{183}-\eta_{27}-\eta_{57}+\eta_{153} \\ 
\eta_{9}-\eta_{201}-\eta_{51}+\eta_{81} & 
\eta_{171}-\eta_{39}-\eta_{129}+\eta_{81} & 
\eta_{69}-\eta_{141}-\eta_{111}+\eta_{99} 
\end{array} \right), 
\nonumber \\ 
y_{ij}^3 &=& \frac{1}{\sqrt{2}} 
\left( \begin{array}{ccc}
\sqrt{2}(\eta_{100}-\eta_{140}) & 
\sqrt{2}(\eta_{80}-\eta_{200}) & 
\sqrt{2}(\eta_{160}-\eta_{20}) \\
\eta_{68}-\eta_{208}-\eta_{128}+\eta_{152} & 
\eta_{172}-\eta_{32}-\eta_{52}+\eta_{88} & 
\eta_{8}-\eta_{148}-\eta_{188}+\eta_{92} \\ 
\eta_{184}-\eta_{44}-\eta_{124}+\eta_{164} & 
\eta_{4}-\eta_{136}-\eta_{116}+\eta_{164} & 
\eta_{176}-\eta_{104}-\eta_{64}+\eta_{76} 
\end{array} \right), 
\nonumber \\
y_{ij}^4 &=& \frac{1}{\sqrt{2}} 
\left( \begin{array}{ccc}
\sqrt{2}(\eta_{145}-\eta_{205}) & 
\sqrt{2}(\eta_{95}-\eta_{25}) & 
\sqrt{2}(\eta_{85}-\eta_{155}) \\
\eta_{107}-\eta_{37}-\eta_{47}+\eta_{23} & 
\eta_{73}-\eta_{143}-\eta_{193}+\eta_{157} & 
\eta_{167}-\eta_{97}-\eta_{13}+\eta_{83} \\ 
\eta_{61}-\eta_{131}-\eta_{121}+\eta_{11} & 
\eta_{179}-\eta_{109}-\eta_{59}+\eta_{11} & 
\eta_{1}-\eta_{71}-\eta_{181}+\eta_{169} 
\end{array} \right), 
\nonumber 
\end{eqnarray}
in the short notation $\eta_N$ defined in Eq.~(\ref{eq:notation-y}) with 
$M=M_1M_2M_3=420$.

\subsubsection{5-8-13 model}

Here we show the model with 
$(|I^{(1)}_{ab}|,|I^{(1)}_{ca}|,|I^{(1)}_{bc}|)=(5,8,13)$.
The following table shows zero-mode wavefunctions of left, right-handed 
matter fields and Higgs fields.
\begin{center}
\begin{tabular}{c|c|c|c}
& $L_i (\lambda^{ab})$ & $R_j (\lambda^{ca})$ & $H_k (\lambda^{bc})$ 
\\ \hline
0 & 
$\Theta^{0,5}$ & $\frac{1}{\sqrt{2}}(\Theta^{1,8}-\Theta^{7,8})$ & 
$\frac{1}{\sqrt{2}}(\Theta^{1,13}-\Theta^{12,13})$ \\  
1 & 
$\frac{1}{\sqrt{2}}\left(\Theta^{1,5}+\Theta^{4,5}\right)$ & 
$\frac{1}{\sqrt{2}}\left(\Theta^{2,8}-\Theta^{6,8}\right)$ & 
$\frac{1}{\sqrt{2}}\left(\Theta^{2,13}-\Theta^{11,13}\right)$ \\  
2 & 
$\frac{1}{\sqrt{2}}\left(\Theta^{2,5}+\Theta^{3,5}\right)$ &
$\frac{1}{\sqrt{2}}\left(\Theta^{3,8}-\Theta^{5,8}\right)$ & 
$\frac{1}{\sqrt{2}}\left(\Theta^{3,13}-\Theta^{10,13}\right)$ \\  
3 & - & - & 
$\frac{1}{\sqrt{2}}\left(\Theta^{4,13}-\Theta^{9,13}\right)$ \\  
4 & - & - & 
$\frac{1}{\sqrt{2}}\left(\Theta^{5,13}-\Theta^{8,13}\right)$ \\  
5 & - & - & 
$\frac{1}{\sqrt{2}}\left(\Theta^{6,13}-\Theta^{7,13}\right)$ \\  
\end{tabular}
\end{center}
This model has six zero-modes for the Higgs fields.
Yukawa couplings $Y_{ijk} L_i R_j H_k$ are given by 
\begin{eqnarray}
Y_{ij}^k H_k &=& 
  y_{ij}^0 H_0 + y_{ij}^1 H_1 + y_{ij}^2 H_2 
+ y_{ij}^3 H_3 + y_{ij}^4 H_4 + y_{ij}^5 H_5, 
\nonumber
\end{eqnarray}
where 
\begin{eqnarray}
y_{ij}^0 &=& \frac{1}{\sqrt{2}} 
\left( \begin{array}{ccc}
\sqrt{2}(\eta_5-\eta_{125}) & 
\sqrt{2}(\eta_{190}-\eta_{70}) & 
\sqrt{2}(\eta_{135}-\eta_{255}) \\
\eta_{203}+\eta_{213}-\eta_{83}-\eta_{187} & 
\eta_{122}-\eta_{138}+\eta_{18}-\eta_{242} & 
\eta_{73}-\eta_{57}+\eta_{177}-\eta_{47} \\ 
\eta_{109}-\eta_{21}+\eta_{99}-\eta_{229} & 
\eta_{86}-\eta_{174}+\eta_{226}-\eta_{34} & 
\eta_{239}-\eta_{151}+\eta_{31}-\eta_{161} 
\end{array} \right), 
\nonumber \\ 
y_{ij}^1 &=& \frac{1}{\sqrt{2}} 
\left( \begin{array}{ccc}
\sqrt{2}(\eta_{205}-\eta_{75}) & 
\sqrt{2}(\eta_{10}-\eta_{250}) & 
\sqrt{2}(\eta_{185}-\eta_{55}) \\
\eta_{3}+\eta_{237}-\eta_{133}-\eta_{107} & 
\eta_{198}-\eta_{62}+\eta_{132}-\eta_{152} & 
\eta_{127}-\eta_{257}+\eta_{23}-\eta_{153} \\ 
\eta_{211}-\eta_{179}+\eta_{101}-\eta_{29} & 
\eta_{114}-\eta_{146}+\eta_{94}-\eta_{166} & 
\eta_{81}-\eta_{49}+\eta_{231}-\eta_{159} 
\end{array} \right), 
\nonumber \\
y_{ij}^2 &=& \frac{1}{\sqrt{2}} 
\left( \begin{array}{ccc}
\sqrt{2}(\eta_{115}-\eta_{245}) & 
\sqrt{2}(\eta_{210}-\eta_{50}) & 
\sqrt{2}(\eta_{15}-\eta_{145}) \\
\eta_{197}+\eta_{137}-\eta_{67}-\eta_{93} & 
\eta_{2}-\eta_{258}+\eta_{102}-\eta_{158} & 
\eta_{193}-\eta_{63}+\eta_{223}-\eta_{167} \\ 
\eta_{11}-\eta_{141}+\eta_{219}-\eta_{171} & 
\eta_{206}-\eta_{54}+\eta_{106}-\eta_{154} & 
\eta_{119}-\eta_{249}+\eta_{89}-\eta_{41} 
\end{array} \right), 
\nonumber \\ 
y_{ij}^3 &=& \frac{1}{\sqrt{2}} 
\left( \begin{array}{ccc}
\sqrt{2}(\eta_{85}-\eta_{45}) & 
\sqrt{2}(\eta_{110}-\eta_{150}) & 
\sqrt{2}(\eta_{135}-\eta_{255}) \\
\eta_{123}+\eta_{163}-\eta_{253}-\eta_{227} & 
\eta_{202}-\eta_{58}+\eta_{98}-\eta_{162} & 
\eta_{7}-\eta_{137}+\eta_{97}-\eta_{33} \\ 
\eta_{189}-\eta_{59}+\eta_{19}-\eta_{149} & 
\eta_{6}-\eta_{254}+\eta_{214}-\eta_{46} & 
\eta_{201}-\eta_{71}+\eta_{111}-\eta_{241} 
\end{array} \right), 
\nonumber \\
y_{ij}^4 &=& \frac{1}{\sqrt{2}} 
\left( \begin{array}{ccc}
\sqrt{2}(\eta_{235}-\eta_{155}) & 
\sqrt{2}(\eta_{90}-\eta_{170}) & 
\sqrt{2}(\eta_{105}-\eta_{25}) \\
\eta_{77}+\eta_{157}-\eta_{53}-\eta_{27} & 
\eta_{118}-\eta_{142}+\eta_{222}-\eta_{38} & 
\eta_{207}-\eta_{183}+\eta_{103}-\eta_{233} \\ 
\eta_{131}-\eta_{259}+\eta_{181}-\eta_{51} & 
\eta_{194}-\eta_{66}+\eta_{14}-\eta_{246} & 
\eta_{1}-\eta_{129}+\eta_{209}-\eta_{79} 
\end{array} \right), 
\nonumber \\
y_{ij}^5 &=& \frac{1}{\sqrt{2}} 
\left( \begin{array}{ccc}
\sqrt{2}(\eta_{35}-\eta_{165}) & 
\sqrt{2}(\eta_{230}-\eta_{30}) & 
\sqrt{2}(\eta_{95}-\eta_{225}) \\
\eta_{243}+\eta_{43}-\eta_{147}-\eta_{173} & 
\eta_{82}-\eta_{178}+\eta_{22}-\eta_{238} & 
\eta_{113}-\eta_{17}+\eta_{217}-\eta_{87} \\ 
\eta_{69}-\eta_{61}+\eta_{139}-\eta_{251} & 
\eta_{126}-\eta_{134}+\eta_{186}-\eta_{74} & 
\eta_{199}-\eta_{191}+\eta_{9}-\eta_{121} 
\end{array} \right), 
\nonumber 
\end{eqnarray}
in the short notation $\eta_N$ defined in Eq.~(\ref{eq:notation-y}) with 
$M=M_1M_2M_3=520$.

\subsubsection{5-8-3 model}

Here we show the model with 
$(|I^{(1)}_{ab}|,|I^{(1)}_{ca}|,|I^{(1)}_{bc}|)=(5,8,3)$.
The following table shows zero-mode wavefunctions of left, right-handed 
matter fields and Higgs fields.
\begin{center}
\begin{tabular}{c|c|c|c}
& $L_i (\lambda^{ab})$ & $R_j (\lambda^{ca})$ & $H_k (\lambda^{bc})$ 
\\ \hline
0 & 
$\Theta^{0,5}$ & $\frac{1}{\sqrt{2}}(\Theta^{1,8}-\Theta^{7,8})$ 
& $\frac{1}{\sqrt{2}}(\Theta^{1,3}-\Theta^{2,3})$ \\  
1 & 
$\frac{1}{\sqrt{2}}\left(\Theta^{1,5}+\Theta^{4,5}\right)$ & 
$\frac{1}{\sqrt{2}}\left(\Theta^{2,8}-\Theta^{6,8}\right)$ & 
- \\  
2 & 
$\frac{1}{\sqrt{2}}(\Theta^{2,5}+\Theta^{3,5}) $ & 
$\frac{1}{\sqrt{2}}\left(\Theta^{3,8}-\Theta^{5,8}\right)$ & 
- \\  
\end{tabular}
\end{center}
This model has a single zero-mode for the Higgs field.
Yukawa couplings $Y_{ijk} L_i R_j H_k$ are given by 
\begin{eqnarray}
Y_{ij}^k H_k &=& \frac{1}{\sqrt{2}} 
\left( \begin{array}{ccc}
\sqrt{2}(\eta_5-\eta_{35}) & 
\sqrt{2}(\eta_{50}-\eta_{10}) & 
\sqrt{2}(\eta_{25}-\eta_{55}) \\
\eta_{43} -\eta_{37}-\eta_{13}+\eta_{53} & 
\eta_{2} -\eta_{38}-\eta_{58}+\eta_{22}  &
\eta_{47} -\eta_{7}-\eta_{17}+\eta_{23} \\
\eta_{29} -\eta_{11}-\eta_{59}+\eta_{19} &
\eta_{46} -\eta_{34}-\eta_{14}+\eta_{26} &
\eta_{1} -\eta_{41}-\eta_{31}+\eta_{49} 
\end{array} \right), 
\nonumber   
\end{eqnarray}
in the short notation $\eta_N$ defined in Eq.~(\ref{eq:notation-y}) with 
$M=M_1M_2M_3=120$.

\subsection{(Odd-Odd-Even) wavefunctions}

Here, we study the patterns of Yukawa matrices 
in the models, where zero-modes of left, right-handed 
matter fields and Higgs fields correspond to 
odd, odd and even functions, respectively.

\subsubsection{7-7-14 model}

Here we show the model with 
$(|I^{(1)}_{ab}|,|I^{(1)}_{ca}|,|I^{(1)}_{bc}|)=(7,7,14)$. 
This model is studied in the subsections~\ref{ssec:7-7-14} 
and \ref{ssec:numerical} in detail. 
The zero-mode wavefunctions of left, right-handed 
matter fields and Higgs fields are shown in Table~\ref{7-7-14model}. 

This model has eight zero-modes for the Higgs fields.
Yukawa couplings $Y_{ijk}L_j R_jH_k$ are obtained as 
\begin{eqnarray}
Y_{ijk}H_k &=& 
  y_{ij}^0 H_0 + y_{ij}^1 H_1 + y_{ij}^2 H_2 
+ y_{ij}^3 H_3 + y_{ij}^4 H_4 + y_{ij}^5 H_5
+ y_{ij}^6 H_6 + y_{ij}^7 H_7, 
\nonumber
\end{eqnarray}
where $y_{ij}^k$ is shown in Eq.~(\ref{eq:yij}) 
with $M=M_1M_2M_3=686$.

\subsubsection{7-8-15 model}

Here we show the model with 
$(|I^{(1)}_{ab}|,|I^{(1)}_{ca}|,|I^{(1)}_{bc}|)=(7,8,15)$.
The following table shows zero-mode wavefunctions of left, right-handed 
matter fields and Higgs fields.
\begin{center}
\begin{tabular}{c|c|c|c}
& $L_i (\lambda^{ab})$ & $R_j (\lambda^{ca})$ & $H_k (\lambda^{bc})$ 
\\ \hline
0 & 
$\frac{1}{\sqrt{2}}\left(\Theta^{1,7}-\Theta^{6,7}\right)$ & 
$\frac{1}{\sqrt{2}}\left(\Theta^{1,8}-\Theta^{7,8}\right)$ & 
$\Theta^{0,15}$ \\  
1 & 
$\frac{1}{\sqrt{2}}\left(\Theta^{2,7}-\Theta^{5,7}\right)$ & 
$\frac{1}{\sqrt{2}}\left(\Theta^{2,8}-\Theta^{6,8}\right)$ & 
$\frac{1}{\sqrt{2}}\left(\Theta^{1,15}+\Theta^{14,15}\right)$ \\  
2 & 
$\frac{1}{\sqrt{2}}\left(\Theta^{3,7}-\Theta^{4,7}\right)$ & 
$\frac{1}{\sqrt{2}}\left(\Theta^{3,7}-\Theta^{5,8}\right)$ & 
$\frac{1}{\sqrt{2}}\left(\Theta^{2,15}+\Theta^{13,15}\right)$ \\  
3 & - & - & 
$\frac{1}{\sqrt{2}}\left(\Theta^{3,15}+\Theta^{12,15}\right)$ \\  
4 & - & - & 
$\frac{1}{\sqrt{2}}\left(\Theta^{4,15}+\Theta^{11,15}\right)$ \\  
5 & - & - & 
$\frac{1}{\sqrt{2}}\left(\Theta^{5,15}+\Theta^{10,15}\right)$ \\  
6 & - & - & 
$\frac{1}{\sqrt{2}}\left(\Theta^{6,15}+\Theta^{9,15}\right)$ \\  
7 & - & - & 
$\frac{1}{\sqrt{2}}\left(\Theta^{7,15}+\Theta^{8,15}\right)$ \\  
\end{tabular}
\end{center}
This model has eight zero-modes for the Higgs fields.
Yukawa couplings $Y_{ijk} L_i R_j H_k$ are given by
\begin{eqnarray}
Y_{ij}^k H_k &=& 
  y_{ij}^0 H_0 + y_{ij}^1 H_1 + y_{ij}^2 H_2 
+ y_{ij}^3 H_3 + y_{ij}^4 H_4 + y_{ij}^5 H_5
+ y_{ij}^6 H_6 + y_{ij}^7 H_7, 
\nonumber
\end{eqnarray}
where 
\begin{eqnarray}
y_{ij}^0 &=& 
\left( \begin{array}{ccc}
\eta_{225}-\eta_{15} & 
\eta_{330}-\eta_{90} &
\eta_{405}-\eta_{195} \\
\eta_{345}-\eta_{135} & 
\eta_{390}-\eta_{30} &
\eta_{285}-\eta_{75} \\
\eta_{375}-\eta_{255} & 
\eta_{270}-\eta_{150} &
\eta_{165}-\eta_{45} \\
\end{array} \right), 
\nonumber \\
y_{ij}^1 &=& \frac{1}{\sqrt{2}} 
\left( \begin{array}{ccc}
\eta_{113}-\eta_{97}-\eta_{127}+\eta_{337} & 
\eta_{218}-\eta_{202}-\eta_{22}+\eta_{398} & 
\eta_{323}-\eta_{307}-\eta_{83}+\eta_{293} \\
\eta_{233}-\eta_{23}-\eta_{247}+\eta_{383} &
\eta_{338}-\eta_{82}-\eta_{142}+\eta_{278} &
\eta_{397}-\eta_{187}-\eta_{37}+\eta_{173} \\
\eta_{353}-\eta_{143}-\eta_{367}+\eta_{263} &
\eta_{382}-\eta_{38}-\eta_{262}+\eta_{158} &
\eta_{277}-\eta_{67}-\eta_{157}+\eta_{53} \\
\end{array} \right), 
\nonumber \\
y_{ij}^2 &=& \frac{1}{\sqrt{2}} 
\left( \begin{array}{ccc}
\eta_{1}-\eta_{209}-\eta_{239}+\eta_{391} &
\eta_{106}-\eta_{314}-\eta_{134}+\eta_{286} &
\eta_{211}-\eta_{419}-\eta_{29}+\eta_{181} \\
\eta_{121}-\eta_{89}-\eta_{359}+\eta_{271} &
\eta_{226}-\eta_{194}-\eta_{254}+\eta_{166} &
\eta_{331}-\eta_{299}-\eta_{149}+\eta_{61} \\
\eta_{241}-\eta_{31}-\eta_{361}+\eta_{151} &
\eta_{346}-\eta_{74}-\eta_{374}+\eta_{46} &
\eta_{389}-\eta_{179}-\eta_{269}+\eta_{59} \\
\end{array} \right), 
\nonumber \\
y_{ij}^3 &=& \frac{1}{\sqrt{2}} 
\left( \begin{array}{ccc}
\eta_{111}-\eta_{321}-\eta_{351}+\eta_{279} &
\eta_{6}-\eta_{414}-\eta_{246}+\eta_{174} &
\eta_{99}-\eta_{309}-\eta_{141}+\eta_{69} \\
\eta_{9}-\eta_{201}-\eta_{369}+\eta_{159} &
\eta_{114}-\eta_{306}-\eta_{366}+\eta_{54} &
\eta_{219}-\eta_{411}-\eta_{261}+\eta_{51} \\
\eta_{129}-\eta_{81}-\eta_{249}+\eta_{39} &
\eta_{234}-\eta_{186}-\eta_{354}+\eta_{66} &
\eta_{339}-\eta_{291}-\eta_{381}+\eta_{171} \\
\end{array} \right), 
\nonumber \\
y_{ij}^4 &=& \frac{1}{\sqrt{2}} 
\left( \begin{array}{ccc}
\eta_{223}-\eta_{407}-\eta_{377}+\eta_{167} &
\eta_{118}-\eta_{302}-\eta_{358}+\eta_{62} &
\eta_{13}-\eta_{197}-\eta_{253}+\eta_{43} \\
\eta_{103}-\eta_{313}-\eta_{257}+\eta_{47} &
\eta_{2}-\eta_{418}-\eta_{362}+\eta_{58} &
\eta_{107}-\eta_{317}-\eta_{373}+\eta_{163} \\
\eta_{17}-\eta_{193}-\eta_{137}+\eta_{73} &
\eta_{122}-\eta_{298}-\eta_{242}+\eta_{178} &
\eta_{227}-\eta_{403}-\eta_{347}+\eta_{283} \\
\end{array} \right), 
\nonumber \\
y_{ij}^5 &=& \frac{1}{\sqrt{2}} 
\left( \begin{array}{ccc}
\eta_{335}-\eta_{295}-\eta_{265}+\eta_{55} &
\eta_{230}-\eta_{190}-\eta_{370}+\eta_{50} &
\eta_{125}-\eta_{85}-\eta_{365}+\eta_{155} \\
\eta_{215}-\eta_{415}-\eta_{145}+\eta_{65} &
\eta_{110}-\eta_{310}-\eta_{250}+\eta_{170} &
\eta_{5}-\eta_{205}-\eta_{355}+\eta_{275} \\
\eta_{95}-\eta_{305}-\eta_{25}+\eta_{185} &
\eta_{10}-\eta_{410}-\eta_{130}+\eta_{290} &
\eta_{115}-\eta_{325}-\eta_{235}+\eta_{395} \\
\end{array} \right), 
\nonumber \\
y_{ij}^6 &=& \frac{1}{\sqrt{2}} 
\left( \begin{array}{ccc}
\eta_{393}-\eta_{183}-\eta_{153}+\eta_{57} &
\eta_{342}-\eta_{78}-\eta_{258}+\eta_{162} &
\eta_{237}-\eta_{27}-\eta_{363}+\eta_{267} \\
\eta_{327}-\eta_{303}-\eta_{33}+\eta_{177} &
\eta_{222}-\eta_{198}-\eta_{138}+\eta_{282} &
\eta_{117}-\eta_{93}-\eta_{243}+\eta_{387} \\
\eta_{207}-\eta_{417}-\eta_{87}+\eta_{297} &
\eta_{102}-\eta_{318}-\eta_{18}+\eta_{402} &
\eta_{3}-\eta_{213}-\eta_{123}+\eta_{333} \\
\end{array} \right), 
\nonumber \\
y_{ij}^7 &=& \frac{1}{\sqrt{2}} 
\left( \begin{array}{ccc}
\eta_{281}-\eta_{71}-\eta_{41}+\eta_{169} &
\eta_{386}-\eta_{34}-\eta_{146}+\eta_{274} &
\eta_{349}-\eta_{139}-\eta_{251}+\eta_{379} \\
\eta_{401}-\eta_{191}-\eta_{79}+\eta_{289} &
\eta_{334}-\eta_{86}-\eta_{26}+\eta_{394} &
\eta_{229}-\eta_{19}-\eta_{131}+\eta_{341} \\
\eta_{319}-\eta_{311}-\eta_{199}+\eta_{409} &
\eta_{214}-\eta_{206}-\eta_{94}+\eta_{326} &
\eta_{109}-\eta_{101}-\eta_{11}+\eta_{221} \\
\end{array} \right), 
\nonumber
\end{eqnarray}
in the short notation $\eta_N$ defined in Eq.~(\ref{eq:notation-y}) with 
$M=M_1M_2M_3=840$.

\subsubsection{7-8-1 model}

Here we show the model with 
$(|I^{(1)}_{ab}|,|I^{(1)}_{ca}|,|I^{(1)}_{bc}|)=(7,8,1)$.
The following table shows zero-mode wavefunctions of left, right-handed 
matter fields and Higgs fields.
\begin{center}
\begin{tabular}{c|c|c|c}
& $L_i (\lambda^{ab})$ & $R_j (\lambda^{ca})$ & $H_k (\lambda^{bc})$ 
\\ \hline
0 & 
$\frac{1}{\sqrt{2}}(\Theta^{1,7}-\Theta^{6,7})$ & 
$\frac{1}{\sqrt{2}}(\Theta^{1,8}-\Theta^{7,8})$ 
& $\Theta^{0,1}$ \\  
1 & 
$\frac{1}{\sqrt{2}}\left(\Theta^{2,7}-\Theta^{5,7}\right)$ & 
$\frac{1}{\sqrt{2}}\left(\Theta^{2,8}-\Theta^{6,8}\right)$ & 
- \\  
2 & 
$\frac{1}{\sqrt{2}}\left(\Theta^{3,7}-\Theta^{4,7}\right)$ & 
$\frac{1}{\sqrt{2}}\left(\Theta^{3,8}-\Theta^{5,8}\right)$ & 
- \\  
\end{tabular}
\end{center}
This model has a single zero-mode for the Higgs field.
Yukawa couplings $Y_{ijk} L_i R_j H_k$ are given by
\begin{eqnarray}
Y_{ij}^k H_k &=& \frac{1}{\sqrt{2}} H_0 
\left( \begin{array}{ccc}
\sqrt{2}(\eta_5-\eta_{35}) & 
\sqrt{2}(\eta_{50}-\eta_{10}) & 
\sqrt{2}(\eta_{25}-\eta_{55}) \\
\eta_{43} -\eta_{37}-\eta_{13}+\eta_{53} & 
\eta_{2} -\eta_{38}-\eta_{58}+\eta_{22} & 
\eta_{47} -\eta_{7}-\eta_{17}+\eta_{23} \\
\eta_{29} -\eta_{11}-\eta_{59}+\eta_{19} & 
\eta_{46} -\eta_{34}-\eta_{14}+\eta_{26} & 
\eta_{1} -\eta_{41}-\eta_{31}+\eta_{49} \\
\end{array} \right), \nonumber   
\end{eqnarray}
in the short notation $\eta_N$ defined in Eq.~(\ref{eq:notation-y}) with 
$M=M_1M_2M_3=56$.

\subsubsection{8-8-16 model}

Here we show the model with 
$(|I^{(1)}_{ab}|,|I^{(1)}_{ca}|,|I^{(1)}_{bc}|)=(8,8,16)$.
The following table shows zero-mode wavefunctions of left, right-handed 
matter fields and Higgs fields.
\begin{center}
\begin{tabular}{c|c|c|c}
& $L_i (\lambda^{ab})$ & $R_j (\lambda^{ca})$ & $H_k (\lambda^{bc})$ 
\\ \hline
0 & 
$\frac{1}{\sqrt{2}}\left(\Theta^{1,8}-\Theta^{7,8}\right)$ & 
$\frac{1}{\sqrt{2}}\left(\Theta^{1,8}-\Theta^{7,8}\right)$ & 
$\Theta^{0,16}$ \\  
1 & 
$\frac{1}{\sqrt{2}}\left(\Theta^{2,8}-\Theta^{6,8}\right)$ & 
$\frac{1}{\sqrt{2}}\left(\Theta^{2,8}-\Theta^{6,8}\right)$ & 
$\frac{1}{\sqrt{2}}\left(\Theta^{1,16}+\Theta^{15,16}\right)$ \\  
2 & 
$\frac{1}{\sqrt{2}}\left(\Theta^{3,8}-\Theta^{5,8}\right)$ & 
$\frac{1}{\sqrt{2}}\left(\Theta^{3,8}-\Theta^{5,8}\right)$ & 
$\frac{1}{\sqrt{2}}\left(\Theta^{2,16}+\Theta^{14,16}\right)$ \\  
3 & - & - & 
$\frac{1}{\sqrt{2}}\left(\Theta^{3,16}+\Theta^{13,16}\right)$ \\  
4 & - & - & 
$\frac{1}{\sqrt{2}}\left(\Theta^{4,16}+\Theta^{12,16}\right)$ \\  
5 & - & - & 
$\frac{1}{\sqrt{2}}\left(\Theta^{5,16}+\Theta^{11,16}\right)$ \\  
6 & - & - & 
$\frac{1}{\sqrt{2}}\left(\Theta^{6,16}+\Theta^{10,16}\right)$ \\  
7 & - & - & 
$\frac{1}{\sqrt{2}}\left(\Theta^{7,16}+\Theta^{9,16}\right)$ \\  
8 & - & - & 
$\Theta^{8,16}$ \\  
\end{tabular}
\end{center}
This model has eight zero-modes for the Higgs fields.
Yukawa couplings $Y_{ijk} L_i R_j H_k$ are obtained as 
\begin{eqnarray}
Y_{ij}^k H_k &=& 
  y_{ij}^0 H_0 + y_{ij}^1 H_1 + y_{ij}^2 H_2 
+ y_{ij}^3 H_3 + y_{ij}^4 H_4 + y_{ij}^5 H_5
+ y_{ij}^6 H_6 + y_{ij}^7 H_7 + y_{ij}^8 H_8, 
\nonumber
\end{eqnarray}
where 
\begin{eqnarray}
y_{ij}^0 &=& 
\left( \begin{array}{ccc}
-y_g & 0 & 0 \\
0 & -y_e & 0 \\
0 & 0 & -y_g 
\end{array} \right), \qquad
y_{ij}^1 \ = \ \frac{1}{\sqrt{2}} 
\left( \begin{array}{ccc}
0 & -y_d & 0 \\
-y_d & 0 & -y_f \\
0 & -y_f & 0
\end{array} \right), 
\nonumber \\
y_{ij}^2 &=& \frac{1}{\sqrt{2}} 
\left( \begin{array}{ccc}
y_a & 0 &  -y_e  \\
0 & 0 & 0 \\ 
-y_e & 0 & y_i    
\end{array} \right), \qquad 
y_{ij}^3 \ = \ \frac{1}{\sqrt{2}} 
\left( \begin{array}{ccc}
0 & y_b  &  0 \\
y_b & 0 & y_h \\ 
0 & y_h & 0  
\end{array} \right), 
\nonumber \\
y_{ij}^4 &=& \frac{1}{\sqrt{2}} 
\left( \begin{array}{ccc}
0 & 0 & y_c+y_g \\ 
0 & y_a+y_i & 0 \\ 
y_c+y_g & 0 & 0 
\end{array} \right), \qquad 
y_{ij}^5 \ = \ \frac{1}{\sqrt{2}} 
\left( \begin{array}{ccc}
0 & y_h  & 0 \\
y_h & 0 & y_b \\ 
0 & y_b & 0 
\end{array} \right), 
\nonumber \\
y_{ij}^6 &=& \frac{1}{\sqrt{2}} 
\left( \begin{array}{ccc}
y_i & 0 &  -y_e  \\
0 & 0 & 0 \\ 
-y_e & 0 & y_a 
\end{array} \right), \qquad 
y_{ij}^7 \ = \ \frac{1}{\sqrt{2}} 
\left( \begin{array}{ccc}
0 & -y_f & 0 \\
-y_f & 0 & -y_d \\
0 & -y_d & 0
\end{array} \right), 
\nonumber \\
y_{ij}^8 &=& 
\left( \begin{array}{ccc}
-y_c & 0 & 0 \\
0 & -y_e & 0 \\
0 & 0 & -y_c 
\end{array} \right), 
\nonumber
\end{eqnarray}
and
\begin{eqnarray}
y_a &=& \eta_0 +2(\eta_{128} +2\eta_{256} +2\eta_{384})+\eta_{512}, 
\nonumber \\ 
y_b &=& \eta_{8} + \eta_{120} + \eta_{136} + \eta_{248} + \eta_{264}
          + \eta_{376} + \eta_{392} +\eta_{504}, 
\nonumber \\ 
y_c &=& \eta_{16} + \eta_{112} + \eta_{144} + \eta_{240} + \eta_{272}
          + \eta_{368} + \eta_{400} +\eta_{496}, 
\nonumber \\ 
y_d &=& \eta_{24} + \eta_{104} + \eta_{156} + \eta_{232} + \eta_{280}
          + \eta_{360} + \eta_{408} +\eta_{488}, 
\nonumber \\ 
y_e &=& \eta_{32} + \eta_{96} + \eta_{164} + \eta_{224} + \eta_{288}
          + \eta_{352} + \eta_{416} +\eta_{480}, 
\nonumber \\ 
y_f &=& \eta_{40} + \eta_{88} + \eta_{172} + \eta_{216} + \eta_{296}
          + \eta_{344} + \eta_{424} +\eta_{472}, 
\nonumber \\ 
y_g &=& \eta_{48} + \eta_{80} + \eta_{180} + \eta_{208} + \eta_{304}
          + \eta_{336} + \eta_{432} +\eta_{464}, 
\nonumber \\ 
y_h &=& \eta_{56} + \eta_{72} + \eta_{188} + \eta_{200}, + \eta_{312}
          + \eta_{328} + \eta_{440} +\eta_{456}, 
\nonumber \\
y_i &=& 2(\eta_{64} +\eta_{192} +\eta_{320}+\eta_{448}), 
\nonumber  
\end{eqnarray}
in the short notation $\eta_N$ defined in Eq.~(\ref{eq:notation-y}) with 
$M=M_1M_2M_3=1024$.

\end{document}